\documentclass{article}
\usepackage{graphicx}  
\usepackage{amsmath}   
\usepackage[compress]{cite}
\usepackage{amssymb}   
\usepackage{bm} 
\usepackage{dcolumn}
\usepackage{color}
\usepackage{mathrsfs}
\usepackage{amsfonts}
\usepackage{varioref}
\usepackage{textcomp}
\RequirePackage[colorlinks,citecolor=blue,urlcolor=magenta,linkcolor=blue]{hyperref}
\allowdisplaybreaks
\newcounter{mnotecount}[section]

\renewcommand{\themnotecount}{\thesection.\arabic{mnotecount}}

\newcommand{\mnotex}[1]
{\protect{\stepcounter{mnotecount}}$^{\mbox{\footnotesize
$
\bullet$\themnotecount}}$ \marginpar{
\raggedright\small\em
$\!\!\!\!\!\!\,\bullet$\themnotecount: #1} }

\labelformat{section}{Section #1} 
\labelformat{subsection}{Section #1} 
\labelformat{subsubsection}{Section #1}
\labelformat{subsubsubsection}{Section #1}
\labelformat{equation}{Eq.~(#1)} 
\labelformat{figure}{Fig.~#1} 
\labelformat{subfigure}{Fig.~\thefigure#1} 
\labelformat{table}{Table~#1} 
\labelformat{appendix}{Appendix #1}
\newcommand{\be}{\nopagebreak[3]\begin{equation}}
\newcommand{\ee}{\end{equation}}

\newcommand{\ba}{\nopagebreak[3]\begin{eqnarray}}
\newcommand{\ea}{\end{eqnarray}}

\title{\bf Ergoregion instability and echoes for braneworld black holes: Scalar, electromagnetic and gravitational perturbations}
\author{Ramit Dey\footnote{ramitdey@gmail.com}$~^{1}$, Shauvik Biswas\footnote{intsb6@iacs.res.in}$~^{1}$ and Sumanta Chakraborty\footnote{sumantac.physics@gmail.com} $~^{1}$
\\
$^{1}${\small{School of Physical Sciences}}\\
{\small{Indian Association for the Cultivation of Science, Kolkata-700032, India}}}
\date{ }  
\begin{document}
  
\maketitle
\begin{abstract}
In the context of higher dimensional braneworld scenario, we have argued that the occurrence of horizonless exotic compact objects, as an alternative to classical black holes, are more natural. These exotic compact objects carry a distinctive signature of the higher dimension, namely a tidal charge parameter, related to the size of the extra dimension. Due to the absence of any horizon, rotating exotic compact objects are often unstable because of superradiance. Interestingly, these higher dimensional exotic compact objects are more stable than their four dimensional counterpart, as the presence of the tidal charge reduces the size of the extra dimension and hence results into a stronger gravitational field on the brane. A similar inference is drawn by analysing the static modes associated with these exotic compact objects, irrespective of the nature of the perturbation, i.e., it holds true for scalar, electromagnetic and also  gravitational perturbation. The post-merger ringdown phase of the exotic compact object in the braneworld scenario, which can be described in terms of the quasi-normal modes, holds plethora of information regarding the nature of the higher dimension. In this connection we have discussed the analytical computation of the quasi-normal modes  as well as their numerical estimation for perturbations of arbitrary spin, depicting existence of echoes in the ringdown waveform. As we have demonstrated, the echoes in the ringdown waveform depends explicitly on the tidal charge parameter and hence its future detection can provide constraints on the tidal charge parameter, which in turn will enable us to provide a possible bound on the size of the extra dimension.
\end{abstract}
\newpage
\tableofcontents
\newpage
\section{Introduction and Motivation}

Building a consistent  theory of quantum gravity has been one of the biggest open problem in theoretical physics for last couple of decades or so. Despite numerous efforts, so far it has  proved difficult to probe the laws of physics at the Planck scale by some experiment, thereby making the task of unification of general relativity and quantum theory even more challenging. In recent times it has been suggested that observing gravitational waves emitted by merger of binary black holes \cite{Abbott:2016blz,Abbott:2016nmj,Abbott:2017vtc,TheLIGOScientific:2016pea,TheLIGOScientific:2016src,Abbott:2020niy}, one can possibly get some hints about the presence of quantum gravitational effects in the near horizon region of a black hole \cite{PhysRevLett.116.171101,Barack:2018yly}. It is expected that Planckian  physics in the near horizon region of a black hole might remove or replace the classical black hole horizon by some quantum structure \cite{Almheiri:2012rt,Mathur:2005zp}. This will eventually lead to a horizonless, but extremely compact object, which in the literature is often referred to as exotic compact objects. The phrase `exotic' merely stands for the fact that one should not expect the assertions of classical physics to hold true near the surface of this object. Just as it is possible to construct various different models of such horizonless exotic compact objects (henceforth referred to as ECOs), based on particular details of the quantum effects governing the near horizon physics, such as, firewall \cite{Almheiri:2012rt}, fuzzball \cite{Mathur:2005zp}, quantum black holes \cite{Oshita:2019sat,Dey:2020wzm} and 2-2 holes \cite{Conklin:2017lwb,Holdom:2019bdv}, it is also possible to consider other exotic possibilities such as, gravastars\cite{Mazur:2001fv} and wormholes \cite{PhysRevLett.116.171101,Bueno:2017hyj, Bronnikov:2019sbx} within the realm of classical physics as well. One common feature shared by all these models of ECO is the modification of the boundary condition for an ingoing wave in the near horizon region. For a classical black hole, presence of the horizon as a one way membrane ensures that the wave modes are strictly ingoing at the horizon, but for an ECO, the presence of a physical surface in front of the would be horizon \cite{abedi2018echoes,Maggio:2018ivz,Maggio:2019zyv,Maggio:2020jml} or due to the quantum nature of the horizon \cite{Oshita:2018fqu}, an ingoing wave would get partially reflected back. This would give rise to definitive observational signatures, such as the presence of echoes in the post-merger ringdown signal of a binary black hole coalescence \cite{PhysRevLett.116.171101,PhysRevD.94.084031,Abedi:2016hgu,Oshita:2018fqu,Wang:2019rcf}. Tentative detection of these quasi-periodic echoes in the ringdown signal \cite{Abedi:2016hgu,Wang_2018,Conklin:2017lwb,abedi2018echoes,Conklin:2019fcs,Holdom:2019bdv} has made the subject more promising despite some controversies  \cite{Westerweck:2017hus,Tsang:2019zra,Salemi:2019uea,Uchikata:2019frs}, thus demanding more rigorous theoretical and observational analysis.  

In search for a better theoretical motivation for models of quantum black holes, where a near horizon modification to the classical geometry is expected, black holes in the higher dimensional braneworld scenario seems to be a more natural candidate \cite{Dey:2020lhq} (for other gravitational wave related aspects of braneworld scenario, see \cite{Chakraborty:2017qve,Chakravarti:2018vlt,Chakravarti:2019aup,Rahman:2018oso}). By the adaptation of the AdS/CFT correspondence to the Randall Sundrum II (RS2) braneworld scenario \cite{Randall:1999vf,PerezLorenzana:2005iv,Csaki:2004ay,Sundrum:2005jf,Garriga:1999yh}, it was argued in \cite{Emparan_2000,Emparan:2002px, Anderson_2005,Fabbri_2007} that black holes living on the brane \emph{must} be quantum-corrected. This is based on the fact that solving the five dimensional Einstein's equations (the five-dimensional spacetime is referred to as the \emph{bulk}) is equivalent to solving a projected four dimensional Einstein's equations (the four dimensional hypersurface we live in is referred to as the \emph{brane}) coupled with a CFT living on the brane. Due to the coupling of the CFT stress tensor with gravity, any brane localized black hole (which is a solution of the semi-classical field equations on the brane) would inherit quantum effects because of the back-reaction from the CFT and hence the black hole event horizon will either be removed or, replaced with an apparent horizon. Since, the real-world astrophysical black holes always inherit some spin, it is natural to extend the study of the quantum-corrected braneworld black holes to a more relevant, astrophysical setting by incorporating the effect of rotation as well. Following this motivation, in this paper we consider  models of rotating ECOs in the Braneworld scenario. In particular, we investigate the physics of the rotating ECOs living on the brane, having an exterior geometry described by the Kerr-Newman-like metric with the electric charge replaced by the `tidal' charge parameter inherited from the bulk \cite{Aliev:2005bi,Aliev:2009cg} (for other observational signatures of the tidal charge parameter, see \cite{Banerjee:2017hzw,Banerjee:2019sae,Banerjee:2019nnj,Stuchlik:2008fy}). Therefore, these rotating ECOs with a tidal charge and a reflective quantum membrane in the near horizon region can be conceived as a simplified model to capture the near horizon quantum gravity modifications. 

Unlike black holes, the horizonless ECOs with a non-zero spin can develop a strong instability in the ergoregion, as negative energy states are allowed in the ergosphere and if these modes are not curtained by the event horizon they grow exponentially due to superradiance \cite{Starobinsky:1973aij,Starobinsky:1974,Cardoso:2004nk,Brito:2015oca}. It was shown in \cite{Cardoso:2008kj,Maggio:2018ivz} that allowing for a finite absorption by the membrane placed in front of the would-be-horizon, can curb the superradiant instability. In this context it would be interesting to see whether the presence of higher dimensions, in the guise of tidal charge, can suppress the instability. In addition, we will also discuss the physics of the zero frequency (static) mode of scalar, electromagnetic and gravitational perturbation, which is the primary source behind the superradiant instability for ECOs and shall explore the effect of the tidal charge on these modes. Besides superradiance, the post-merger ringdown waveform, which can be described in terms of superposition of QNMs \cite{Kokkotas:1999bd,Berti:2009kk,Nollert:1999ji,Kanti:2006ua,Kanti:2005xa,Konoplya:2011qq}, also depends on the tidal charge parameter and hence on the existence of the higher dimension. In particular, the characteristic signature of horizonless ECOs, which is repeating echoes in the ringdown waveform, will also be different from in presence of higher dimensions. It is also possible to look for plausible observational signatures of braneworld ECOs/quantum black holes, through the echo time-delay measurement, thereby opening up the possibility of constraining the tidal charge with improved observation of echoes. For this purpose, besides following an analytic approach, we also obtain the ringdown waveform using a template for obtaining the gravitational wave signal adapted to the case of horizonless ECOs, as developed in \cite{Mark:2017dnq,Maggio:2019zyv}. 

The paper is organised as follows: In \ref{spin_brane_metric}, we setup the background spacetime describing a rotating ECO in the braneworld scenario and motivate the near horizon quantum modifications to the classical geometry. The basic equations governing any generic spin-s perturbation required to obtain the QNMs and analyse the superradiant instability is established in \ref{linear_pert}. We further discuss the boundary conditions at the horizon/surface of the ECO and at asymptotic infinity in terms of the Teukolsky radial perturbation variable as well as the Detweiler function. In \ref{superradiance}, the superradiant instability for the braneworld ECO has been analyzed, by determining the amplification factor, using an analytic method and later confirmed our predictions by a numerical analysis. The static modes, which plays a crucial role for the mode instability at the ergoregion, has been studied in \ref{static_mode_ECO} for generic spin-s perturbation. \ref{qnm_analytical} deals with determining the frequencies of the QNMs with the modified near horizon boundary condition using an analytic method of asymptotic solution matching, as well as a numerical method based on a ringdown template. Finally, \ref{discussion} summarizes the key results and associated predictions of our analysis along with possible future directions.

\emph{Notations and Conventions:} We will set the fundamental constants $c=1=\hbar$. The lowercase Roman indices $a,b,c,\ldots$ run over the four-dimensional brane spacetime indices, while the uppercase Roman indices $A,B,C,\ldots$ run over the five-dimensional bulk spacetime indices. We use mostly positive signature convention, with the flat spacetime metric being $\textrm{diag}(-1,+1,+1,+1)$, on the brane.  
\section{Spinning exotic compact objects in the braneworld scenario}
\label{spin_brane_metric}

In this section, we will lay down the basic premise of our analysis, namely the exterior geometry of a spinning ECO in the braneworld scenario along with a brief introduction to the braneworld paradigm itself. Even though the braneworld paradigm was originally motivated by the gauge hierarchy problem in the context of particle physics, it has provided significant insights in various aspects of the gravitational physics as well \cite{Randall:1999vf,Randall:1999ee,Horava:1995qa,Kaloper:1999sm,ArkaniHamed:1998rs,Antoniadis:1998ig}. 
In particular, the Randall-Sundrum model with non-compact extra dimensions (referred to as RS2) and its various other incarnations have played a major role in our understanding of gravitational interactions in the presence of higher spacetime dimensions \cite{Randall:1999vf}. 
In what follows, we will mostly be interested in understanding the existence and properties of rotating ECOs in the braneworld models involving non-compact extra dimensions. 

In the RS2 braneworld scenario considered here, the gravitational field, being universal, can propagate along the extra spatial dimension, but matter fields or any other interactions would be constrained to live on the lower dimensional brane. Thus it is legitimate to ask, what will be the `effective' geometry that an observer living on the brane observe? To be precise, one would like to know whether we can have black holes localized on the brane when matter fields collapses under self-gravity \cite{Chamblin_2000}. Often such brane-localized black holes cannot be extended within the bulk due to certain pathological properties \cite{Bruni_2001}. In addition, it has also been observed that the black hole singularity may extend all the way up to the AdS horizon when one studies the bulk extension of these solutions, leading to classical instabilities \cite{Gregory_1993,Gregory_2000}. These classical instabilities were mostly realized as one considers the static general relativistic black holes on the brane and then tries to extrapolate it to the bulk. However, as realized in \cite{Shiromizu:1999wj,Dadhich:2000am,Harko:2004ui},
 the gravitational field equations on the brane are not the standard four-dimensional Einstein's equations, rather they involve corrections from the bulk geometry. First such solution to the `effective' (or, projected) Einstein's equations were derived in \cite{Dadhich:2000am}, albeit in the context of static and spherically symmetric geometry. This has been subsequently generalized to rotating situations as well \cite{Aliev:2005bi}.
These solutions are very difficult to extend within the bulk spacetime, however in the static case such an extension was attempted in \cite{Chamblin_2001}, where it was possible to determine the bulk geometry numerically to some extent. In addition, it was also possible to connect the length of the extra dimension with the additional hairs appearing in these brane-localized solutions, thereby making the higher dimensional length scale accessible to observations. In the subsequent section, we will describe the `effective' gravitational field equations and the relevant solution of these equations. 

\subsection{Solution to the effective gravitational field equations on the brane}\label{sec_solution_brane}

The effective gravitational field equations on the brane can be obtained as follows: (a) one starts by assuming that the bulk geometry satisfies the bulk Einstein's equations, (b) one introduces the projector $h^{A}_{B}=\delta^{A}_{B}-n^{A}n_{B}$, satisfying $h^{A}_{B}n^{B}=0=h^{A}_{B}n_{A}$, where $n_{A}$ is the unit normal to the brane hypersurface and finally (c) using this projector and the Gauss-Codazzi relations one determines projected bulk Einstein tensor on the brane hypersurface and hence determines the effective gravitational field equations on the brane. As emphasized earlier, this equation differs from the four-dimensional Einstein's equations by several additional terms. In vacuum spacetime, the only surviving term corresponds to the electric part of the bulk Weyl tensor and hence the effective gravitational field equations on the brane read \cite{Shiromizu:1999wj,Dadhich:2000am},
\begin{align}\label{eff_grav_eq}
{}^{(4)}G_{ab}+E_{ab}=0~.
\end{align}
Here ${}^{(4)}G_{ab}$ is the induced Einstein tensor on the 3-brane and $E_{ab}\equiv W_{PQRS}n^{P}e^{Q}_{a}n^{R}e^{S}_{b}$ is the electric part of the bulk Weyl tensor $W_{PQRS}$, where $n_{A}$ is the unit normal and $e^{A}_{a}=(\partial x^{A}/\partial y^{a})$ is another form of the projector with $x^{A}$ and $y^{a}$ being the bulk and the brane coordinates respectively. 

Static, spherically symmetric solutions of the above effective gravitational field equations were first derived in \cite{Dadhich:2000am}, which has been subsequently generalized in various other contexts in \cite{Gregory,Maartens:2001jx,Germani:2001du,Harko:2004ui,Chakraborty:2014xla,Chakraborty:2016ydo}. But most of these solutions inherited one common feature, namely the existence of a `tidal charge' parameter having an opposite sign compared to the electric charge in the case of the Reissner-Nordstr\"{o}m black hole. It turns out that the identical feature survives even when one considers stationary and axi-symmetric solutions of the above effective gravitational field equations. Such a rotating black hole solution in the braneworld scenario can be obtained by using a Kerr-Schild type metric ansatz and then solving the above effective gravitational field equations on the 3-brane. In the usual Boyer-Lindquist coordinate system, the metric describing a rotating black hole with mass $M$ and angular momentum $J\equiv aM$ is given as \cite{Aliev:2005bi,Aliev:2009cg} (setting $G=1$ for convenience),
\begin{align} \label{metric}
ds^{2}&=-\left(1-{2Mr+Q\over \rho^2}\right)dt^2-\left({2a\left(2Mr+Q\right) \sin^2\theta \over \rho^2}\right)d\phi dt
+\left(r^2+a^2+{\left(2Mr+Q\right) a^2\sin^2\theta \over \rho^2}\right)\sin^2\theta d\phi^2
\nonumber
\\
&\hskip 2 cm +{\rho^2 \over \Delta}dr^2+\rho^2 d\theta^2~,
\end{align}
where the unknown functions $\Delta(r)$ and $\rho(r,\theta)$ take the following form,
\begin{align}
\rho^2=r^2+a^2\cos^2\theta~, \quad \Delta=r^2+a^2-2Mr-Q~.
\end{align}
The tidal charge parameter $Q$, originating from the bulk Weyl tensor, appears in the above metric through $\Delta(r)$ and can have both positive and negative values. The case of negative $Q$ corresponds to the case of the Kerr-Newman black hole, while the positive $Q$ scenario acts as the crucial discriminator from the usual electric charge and encapsulates the effect of extra dimensions. This distinction becomes apparent as one determines the location of the horizons associated with \ref{metric}, obtained by setting $\Delta=0$, which yields,
\begin{align}\label{horizon}
r_{\pm}=M \pm \sqrt{M^2-a^2+Q}=M\left[1\pm \sqrt{1-\chi^{2}+\frac{Q}{M^{2}}}\right]~;\quad \chi\equiv \frac{a}{M}~.
\end{align}
As evident, for positive $Q$, the two horizons will coincide in the extremal limit, provided $(a/M)>1$, which is in striking contrast with the standard paradigm for rotating black holes. Just for future references, note that the expressions for angular velocity of the event horizon and the Hawking temperature for the braneworld black hole coincides with that of the Kerr black hole, with $r_{+}$ replaced by the horizon radius expressed above in terms of the tidal charge $Q$, such that,
\begin{align}
\Omega_{+}&=\frac{a}{r_{+}^{2}+a^{2}}=\frac{1}{2M}\left(\frac{\chi}{1+\frac{Q}{2M^{2}}+\sqrt{1-\chi^{2}+\frac{Q}{M^{2}}}}\right)~,
\label{angular_velocity}
\\
T_{+}&=\frac{1}{4\pi}\frac{r_{+}-r_{-}}{(r_{+}^{2}+a^{2})}
=\frac{1}{4\pi M}\left(\frac{\sqrt{1-\chi^{2}+\frac{Q}{M^{2}}}}{1+\frac{Q}{2M^{2}}+\sqrt{1-\chi^{2}+\frac{Q}{M^{2}}}}\right)~.
\label{hawking_temp}
\end{align}
This provides a broad overview of the basic physical properties, which we will require, regarding the rotating solutions in the context of effective gravitational field equations in the braneworld scenario. In what follows we will critically review the interpretation of these solutions as black holes and shall argue that these solutions should better be considered as ECOs.

\subsection{Exotic compact objects on the brane}

The above section provides the solution of the effective gravitational field equations on the brane, with stationarity and axi-symmetric configuration. From the perspective of a brane-localized observer, the above solution indeed resembles a black hole, with its horizon located at $r_{+}=M+\sqrt{M^2-a^2+Q}$. However, as we will depict below, in the context of braneworld geometry described above, it is more natural to interpret the rotating solution as an ECO, rather than a black hole.  

For the above surface, located at $r=r_{+}$, to depict an event horizon it is important to consider the global structure of the spacetime, i.e., extension of the horizon to the bulk geometry. However, the extension of the brane-localized geometry, derived above, to the bulk is non-trivial and requires the momentum and the Hamiltonian constraint to be satisfied throughout the bulk extension. Satisfying these constraints throughout the extension of the brane-localized solution into the bulk is difficult and due to growing numerical inaccuracy the extension cannot be performed beyond the AdS radius $\ell$, which is related to the bulk cosmological constant $\Lambda_{5}$ through the relation $\ell \sim \sqrt{-(6/8\pi G_{5}\Lambda_{5})}$ \cite{Randall:1999vf,Shiromizu:1999wj,Chamblin_2001}. This analysis explicitly demonstrates that the surface located at $r=r_{+}$ is an apparent horizon, as this is the outermost surface of the negative expansion of the outgoing null congruences \cite{Chamblin_2001}. This suggests that the brane-localized geometry described by \ref{metric} has an apparent horizon rather than an event horizon and this can be regarded as one of the main feature describing an ECO. Further evidence for the claim that brane-localized geometry, described in \ref{metric}, actually depicts an ECO can be understood from the perspective of AdS/CFT as we elaborate below. 

Since the bulk geometry involves a negative cosmological constant, it is natural to adapt the AdS/CFT correspondence to the braneworld scenario. In the RS2 model it is assumed that our universe is a hypersurface in a $AdS_5$ bulk. According to the AdS/CFT conjecture, the boundary theory of an $AdS$ bulk is a CFT. As a result of this, the zero mode of the five dimensional bulk gravity gets trapped on the brane, inducing a four dimensional gravity coupled to a cut-off CFT. 
The cut-off CFT living on the brane couples to the effective gravitational field equations by introducing the renormalized stress-energy tensor  on the right hand side of \ref{eff_grav_eq}. Therefore, the black holes living on the brane would be quantum corrected due to the back-reaction of the CFT on the brane \cite{Emparan:2002px,Emparan_2000,Emparan_2003}. As a consequence of this, the classical event horizon of the black hole will either be removed or modified due to semi-classical effects of the CFT living on the brane \cite{Anderson_2005,Berthiere:2017tms}. It is expected that any such modifications to the black hole horizon must take place a few Planck length away from the position of the classical horizon and would provide a compelling case for ECOs.  

These results suggest that the brane-localized black hole, introduced above, is actually an ECO, since the horizon is an apparent horizon and there is a CFT living on the brane, leading to quantum corrections to the horizon via back-reaction. Thus following \cite{Dey:2020lhq}, we can replace the event horizon by an apparent horizon and further, as a simplified model of some quantum structure in the near horizon region,  we place a partially reflective membrane a few Planckian distance away from the would be horizon.  As argued in \cite{Dey:2020lhq}, due to a plausible quantum modification to the horizon, the position of the membrane in the braneworld scenario can be related to the size of the extra dimension based on the fact that the apparent horizon of a brane-localized black hole gets shifted by \cite{Fabbri_2007}
\begin{align} \label{rp_shift}
\Delta r_{+}\sim \frac{N^{2}l_{\rm p}^{2}}{M}~;\quad N^{2}\sim \left(\frac{L}{\ell_{\rm p}}\right)^{2}=10^{30}\left(\frac{L}{1~\textrm{mm}}\right)^{2}
\end{align}
where, $N$ corresponds to the degrees of freedom of the CFT living on the brane, $M$ is the mass of the ECO and $l_{\rm p}$ is the four dimensional Planck length. Note that restoring the fundamental constants $G$ and $c$ results into replacing the mass of the black hole in \ref{rp_shift} to $(GM/c^{2})$. This shift in the location of the horizon can also be related to the black hole parameters, by computing the proper length between the horizon and the surface of the ECO. In terms of the Boyer-Lindquist coordinate we can write down the position of the membrane as, $r_{\rm wall}=r_{+}+\Delta r_{+}$, where $\Delta r_{+}$ defines the compactness of the ECO and is comparable to \ref{rp_shift}. Typically, the membrane is assumed to be a Planck proper length away from the horizon, but to be consistent with the semi-classical gravity on the brane we assume that the membrane is placed at some constant, $\eta$, times the Planck proper length $l_{\rm p}$. This implies,
\begin{align}
\int_{r_{+}}^{r_{+}+\Delta r_{+}}\sqrt{g_{rr}}dr\Big|_{\theta=0} \sim \eta l_{\rm p}~.
\end{align}
From this relation we can determine $\Delta r_{+}$, in terms of the black hole parameters as,
\begin{align}\label{epsilon}
\Delta r_{+} \sim {\sqrt{1-\chi^{2}+\frac{Q}{M^{2}}}\over \left(1+\sqrt{1-\chi^{2}+\frac{Q}{M^{2}}}+\frac{Q}{2M^{2}} \right)}\frac{\eta^{2}l_{\rm p}^{2}}{4M}~.
\end{align}
Now, comparing \ref{epsilon} with \ref{rp_shift} one can determine $\eta$ in terms of the black hole parameters and the CFT degrees of freedom $N$, which in turn relates $\eta$ directly to the AdS length scale $L$ as,
\begin{align}
\eta^{2}=4\left(\frac{L}{l_{\rm p}}\right)^{2}\frac{\left(1+\sqrt{1-\chi^{2}+\frac{Q}{M^{2}}}+\frac{Q}{2M^{2}} \right)}{\sqrt{1-\chi^{2}+\frac{Q}{M^{2}}}}~.
\end{align}
Therefore, if one can estimate for the parameter $\eta$, it is possible to get an idea about the AdS length scale associated with the size of the higher dimension. As we will see later, using the computation of QNMs, this estimation of the AdS length scale is indeed possible through time delay measurements. Thus QNMs associated with the perturbation of braneworld ECOs provide a natural ground to observationally probe not only the ECOs but also the higher spatial dimensions. 

\section{Linear perturbations}\label{linear_pert}

Having described the rotating black hole solution in the context of braneworld scenario and the justification behind treating it as the exterior geometry of an exotic compact object, let us concentrate on the linear perturbation of this background geometry. We will consider the case of scalar, electromagnetic and gravitational perturbation, leaving out the perturbation by the Dirac field for a future study. The Klein-Gordon equation, determining the evolution of a scalar field and the Dirac equation, determining the evolution of a Fermionic field are separable in the background metric given by \ref{metric} \cite{page1976dirac,Dadhich:2001sz}. Thus we can employ the techniques developed in the context of Kerr spacetime to the present scenario, in order to find out the  equation for the radial and angular part of the perturbation. However, the electromagnetic and gravitational perturbations are generically non-separable in the background spacetime under consideration. Even though this pose a serious problem, there is a way to get rid of this \cite{dudley1979covariant,PhysRevLett.39.367}. We can either keep the brane geometry fixed and perturb the extra dimensional part, or we can keep the extra dimensional part fixed and perturb the brane geometry. In other words, we consider perturbations which keeps the bulk contribution $E_{ab}$ unchanged but modifies the brane configuration and vice versa. In this case, for reasonable value of the tidal charge $Q$, the electromagnetic and gravitational perturbation also separates in an identical spirit to the Kerr-Newman spacetime\cite{Berti:2005eb}. In what follows, we will assume that such is the case and the perturbation separates out nicely in terms of angular and radial part. This will allow us to discuss various physical properties of the background spacetime arising out of its perturbation. 

\subsection{Basic Equations}

In this section, we will spell out the basic equations governing the linear perturbations of arbitrary spin around the rotating braneworld geometry. These correspond to two different sets of equations, one for the radial part and the other for the  angular part. Under the assumptions mentioned above, the perturbation $\Phi^{(s)}$ associated with a generic spin $s$ is separable, yielding,  
\begin{align}
\Phi^{(s)}=\sum _{\ell,m}e^{i(m\phi-\omega t)}~{}_{s}S_{\ell m}(\theta)~{}_{s}R_{\ell m}(r)~,
\end{align} 
where we have used the result that the spacetime inherits two Killing vector fields, $(\partial/\partial t)^{\mu}$ and $(\partial/\partial \phi)^{\mu}$, respectively. The radial part ${}_{s}R_{\ell m}$ and the angular part ${}_{s}S_{\ell m}$ satisfies the following two equations \cite{Teukolsky:1973ha,Berti:2005eb}, 
\begin{align}
\Delta^{-s}\frac{d}{dr}\bigg(\Delta^{s+1}\dfrac{d\,{}_{s}R_{\ell m}}{dr}\bigg)&+{1\over \Delta}\bigg[K^2-isK{d\Delta\over dr}+\Delta\bigg(2isK'-\lambda \bigg)\bigg] {}_{s}R_{\ell m}=0~,
\label{radial_perturbation}
\\
\frac{d}{dx}\bigg[(1-x^2) \frac{d\,{}_{s}S_{\ell m}}{dx}\bigg]&+\bigg[(a\omega x)^{2}-2a\omega sx+s+{}_{s}A_{\ell m}-{(m+sx)^{2}\over 1-x^2}\bigg]{}_{s}S_{\ell m}=0~,
\label{angular_perturbation}
\end{align}
where we have introduced a new variable, $x\equiv \cos \theta$. In addition, we have introduced the quantity $K\equiv (r^{2}+a^{2})\omega-am$. Also, the separation constants $\lambda$ and ${}_{s}A_{\ell m}$ are related as, $\lambda \equiv {}_{s}A_{\ell m}+a^2\omega^2-2am\omega$. In generic situations, these separation constants need to be computed numerically or using semi-analytical techniques \cite{Berti:2005eb}. We note that for $a=0$ or for $\omega=0$, it is possible to obtain a closed form expression for the separation constants, yielding $\lambda={}_{s}A_{\ell m}=(\ell-s)(\ell+s+1)$. 

The angular equation can be solved, which for scalar perturbation (s=0), yields the associated Legendre polynomial $P_{\ell}^{m}(\cos \theta)$, such that inclusion of $e^{im\phi}$ provides the spherical harmonics $Y_{\ell m}(\theta,\phi)$. For other values of the spin, i.e., for electromagnetic and gravitational perturbation the solution of the angular equation can be expressed in terms of the spin-weighted spherical harmonics. Thus the angular part of the generic spin-s perturbation is well-understood\cite{Berti:2005eb}. 
For the radial part, it is not possible to obtain a closed form expression for the solutions under generic situations. But it is possible to solve this equation at the asymptotic infinity and in the near horizon regime, which we will demonstrate explicitly while solving for the QNMs analytically. The solution of the radial perturbation equation is crucially dependent on the boundary conditions that we impose on the boundaries  i.e., at infinity and on the surface of the ECO. However, the potential for the radial perturbation equation presented above for the Teukolsky variable with non-zero spin, i.e., for electromagnetic and gravitational perturbations, is complex. As a consequence, the asymptotic amplitudes are not simply in terms of exponential functions denoting ingoing and outgoing waves. The way to circumvent this problem is to introduce a new radial function known as the Detweiler function, so that the potential is real and the asymptotic amplitudes are simplified by a transformation of the Teukolsky radial perturbation variable ${}_{s}R_{\ell m}$. In the present context, the Detweiler function can be defined as (following \cite{Detweiler:1977}),
\begin{align}\label{detweiler_fn}
{}_{s}X_{\ell m}=\Delta^{s/2}\sqrt{r^{2}+a^{2}}\left[\alpha(r){}_{s}R_{\ell m}+\beta(r)\Delta^{s+1}\frac{d{}_{s}R_{\ell m}}{dr} \right]~,
\end{align}
where, $\alpha(r)$ and $\beta(r)$ are two unknown radial functions. In terms of this function, the radial perturbation equation presented in \ref{radial_perturbation} takes the following form (for a derivation, see \ref{AppDet}), 
\begin{align}\label{CD_eq}
\frac{d^{2}{}_{s}X_{\ell m}} {dr_{*}^{2}}-V_D(r,\omega){}_{s}X_{\ell m}=0~.
\end{align}
Here, we have introduced the tortoise coordinate $r_{*}$ through the following differential equation, 
\begin{align}\label{tortoise}
\frac{dr_{*}}{dr}=\frac{(r^{2}+a^{2})}{\Delta}~.
\end{align}
The potential $V_D(r,\omega)$ has the following expression,
\begin{align} \label{real_potential}
V_D(r,\omega)&=\frac{U\Delta}{(r^{2}+a^{2})^{2}}+G^{2}+\frac{dG}{dr_{*}}~,
\end{align}
where the functions $G$ and $U$ take the following forms,
\begin{align} \label{G}
G&=\frac{s\Delta'}{2(r^{2}+a^{2})}+\frac{r\Delta}{(r^{2}+a^{2})^{2}}~,
\\
U&=V(r,\omega)+\frac{1}{\beta \Delta^{s}}\left[2\frac{d\alpha}{dr}+\frac{d}{dr}\left(\Delta^{s+1}\frac{d\beta}{dr}\right) \right]~,
\\
V(r,\omega)&=-\frac{1}{\Delta}\left[K^2-isK{d\Delta\over dr}+\Delta\bigg(2isK'-\lambda \bigg)\right]~.
\end{align}
One can show, with appropriate choices of the radial functions $\alpha(r)$ and $\beta(r)$, the potential experienced by the radial part of the spin-s perturbation of rotating ECO in the braneworld scenario is indeed real (for the case of a Kerr-like ECO, see \cite{Maggio:2018ivz}). This is the equation we will use to determine the QNMs as well as superradiant instability associated with the generic spin-s perturbations of the rotating braneworld solution described in \ref{sec_solution_brane}. 

The rest of the paper will be devoted in solving the radial part of the perturbation equation, involving a generic spin-s field, under different boundary conditions appropriate for the physical scenario under consideration. Since the boundary conditions will play a pivotal role in the subsequent analysis, we will discuss them in detail in the subsequent section. 
\subsection{Boundary conditions and reflectivity}\label{reflectivity}

The determination of the QNM spectrum associated with the generic spin-s perturbation of the rotating braneworld solution using the ordinary second order differential equation as presented in \ref{radial_perturbation} requires two boundary conditions. For a classical black hole, one of the boundary condition is imposed at the asymptotic infinity and the other one at the event horizon. As the black hole is a perfect absorber, the perturbation must be \emph{purely} ingoing at the event horizon and at infinity the perturbation must be outgoing. On the other hand, for a horizonless ECO the outgoing boundary condition at infinity remains unchanged while the boundary condition near the surface of the ECO will be modified as the event horizon is either removed or replaced by a partially reflective membrane in front of the would be horizon, due to possibly some quantum effects at least in the present context. 

Let us start by writing down the nature of generic solutions to the radial perturbation equation at infinity and on the horizon for the Teukolsky radial perturbation variable, ${}_{s}R_{\ell m}$. It is advantageous to express the solution in terms of the tortoise coordinate $r_{*}$, defined in \ref{tortoise}, such that for a generic spin $s$ perturbation field we can write down the modes at the asymptotic region $r_{*}\to \pm \infty$ as \cite{Teukolsky:1973ha, Teukolsky:1974yv },
\begin{align} 
{}_{s}R_{\ell m}(r)\sim
    \begin{cases} \label{asymp_ampli}
     \mathcal{I}_{s}r^{-1}e^{-i\omega r_{*}}+\mathcal{O}_{s}r^{-(2s+1)}e^{i\omega r_{*}} & \text{for}\ \quad r_{*}\to \infty
       \\\\
     \mathcal{T}_{s}\Delta^{-s}e^{-i\tilde{\omega} r_{*}}+\mathcal{R}_{s}e^{i\tilde{\omega} r_{*}} & \text{for} \quad r_{*}\to -\infty~.
    \end{cases} 
\end{align}
The horizon frame frequency $\tilde{\omega}$ appearing in the above near horizon behaviour of the Teukolsky radial perturbation variable ${}_{s}R_{\ell m}$ is defined as, $\tilde{\omega}\equiv \omega-m\Omega_{+}$, where $\Omega_{+}$ is the angular velocity of the event horizon, defined in \ref{angular_velocity}. The above asymptotic solutions can be obtained from the asymptotic behaviour of the potential entering the radial perturbation equation presented in \ref{radial_perturbation}. The term with coefficient $\mathcal{I}_{s}$ denotes the ingoing mode while the term with coefficient $\mathcal{O}_{s}$ denotes the outgoing mode at infinity. For the near horizon behaviour, the term with coefficient $\mathcal{T}_{s}$ denotes the ingoing mode and the term with coefficient $\mathcal{R}_{s}$ denotes the outgoing mode. 

\begin{figure}[h]
\begin{center}
\includegraphics[width=10cm]{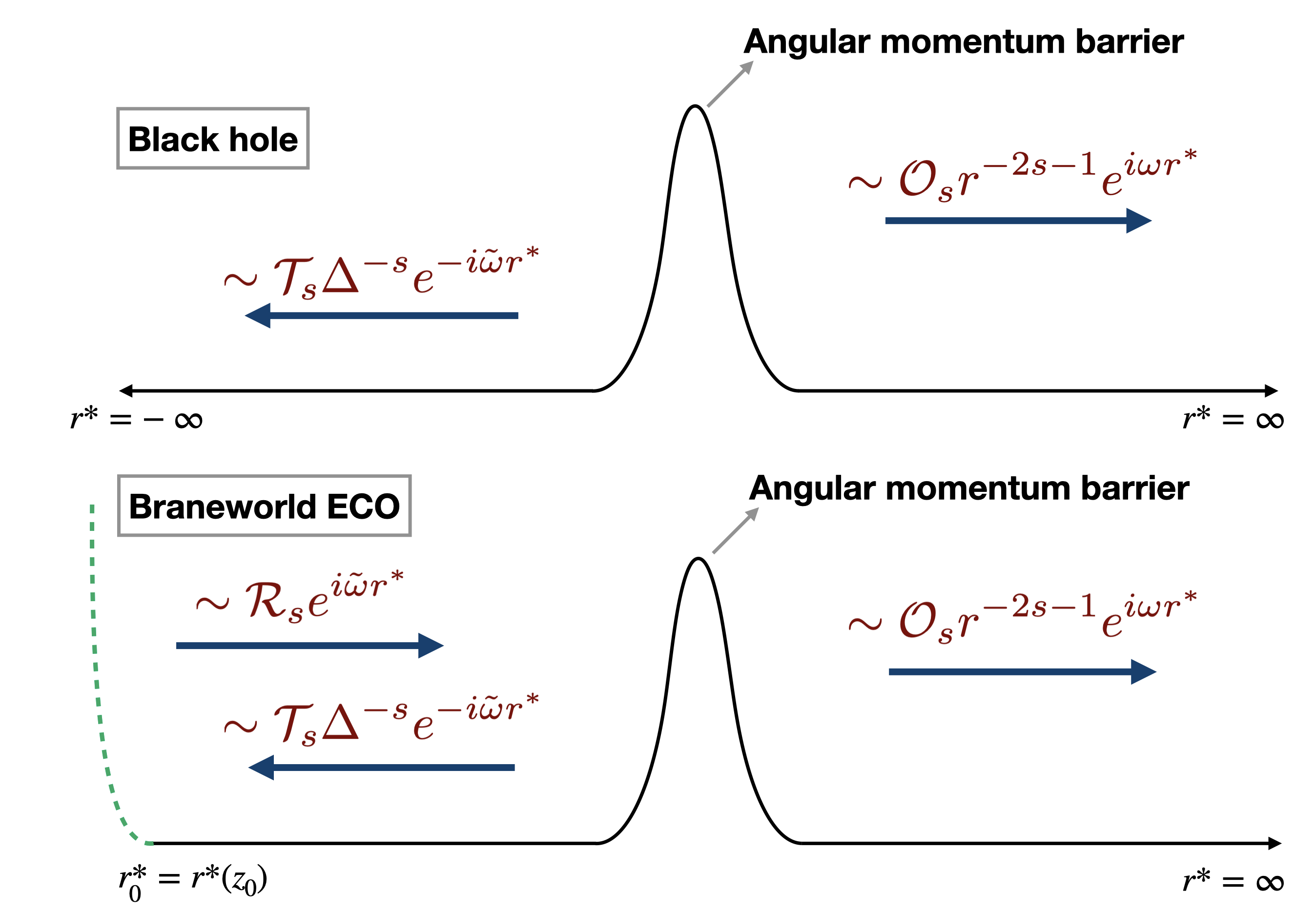}
 \caption[]{The above figure demonstrates the relevant boundary conditions associated with QNMs in the context of a black hole as well as for braneworld ECO. As evident, for black hole spacetime, the perturbation of the angular momentum barrier leads to outgoing mode at infinity and ingoing mode at the horizon. For braneworld ECO, the perturbation leads to outgoing mode at infinity (similar to black hole spacetime) while for the near horizon regime, there will be both ingoing and outgoing modes, in contrast to black hole spacetime. See text for more discussions.}\label{figure:asymp_qnm}
 \end{center}
\end{figure}

In the case of a black hole, the near horizon behaviour of the mode function must be purely ingoing, since nothing can come out of the event horizon. Contrary to the above, in the case of the rotating ECO in the braneworld scenario, the presence of an apparent horizon/quantum structure in the near horizon regime will affect the near horizon boundary condition drastically. Even though the perturbation equations are not directly affected, the solutions will be  different from that of the black hole case due to modification of the near horizon boundary condition. However, the boundary condition at the asymptotic infinity will remain unchanged as shown in \ref{figure:asymp_qnm}. 

In this paper we are mainly concerned with the study of the modes instability developed in the ergosphere of a rotating braneworld ECO due to superradiance and then determine the QNM spectrum of the ECO to obtain its ringdown modes. 
Both, the superradiant instability and the QNM spectrum can be studied/obtained from the same perturbation equation, presented in \ref{radial_perturbation}, but requires different boundary conditions. Let us discuss the boundary conditions associated with the QNMs first, before delving into the corresponding scenario for superradiance. The QNMs are generated by the perturbation of the exterior geometry and hence there will be an outgoing mode at infinity and an ingoing mode to the surface of the ECO. Since the surface of the ECO is reflective in nature, it follows that there will also be an outgoing mode at the near horizon regime and the amplitude of that mode will be proportional to the reflectivity of the surface of the ECO. Thus the appropriate boundary condition for the Teukolsky radial perturbation variable is given by, 
\begin{align} 
{}_{s}R_{\ell m}(r)\sim
    \begin{cases} \label{bound_QNMT}
     \mathcal{O}_{s}r^{-(2s+1)}e^{i\omega r_{*}} & \text{for}\ \quad r_{*}\to \infty
       \\\\
     \mathcal{T}_{s}\Delta^{-s}e^{-i\tilde{\omega} r_{*}}+\mathcal{R}_{s}e^{i\tilde{\omega} r_{*}} & \text{for} \quad r_{*}\to -\infty~.
    \end{cases} 
\end{align}
The physical significance of this boundary condition can also be understood from \ref{figure:asymp_qnm}. In the case of a black hole, perturbation of the angular momentum barrier leads to outgoing modes at infinity and ingoing modes at the horizon. On the other hand in the context of a braneworld ECO, the perturbation of the angular momentum barrier generates outgoing modes at infinity, but in the near-horizon regime we have \emph{both} ingoing and outgoing modes, in stark contrast to the black hole spacetime. 

The quantity $\mathcal{R}_{s}$, which, as we demonstrate, will be related to the reflectivity of the surface of the ECO. The reflectivity of the membrane  placed in front of the would be horizon of the braneworld ECO should ideally depend on the quantum properties of the membrane. Since we considering a generic quantum black hole/ECO in the braneworld scenario and not being specific about a framework describing the quantum origin of the ECO, we have stated the boundary condition at this surface in a generic manner. As we have emphasized earlier, it is often convenient to work with the Detweiler function as it provides a real potential for the perturbation equation, in terms of which the above boundary condition takes the following form, 
\begin{align} \label{echo_sol}
X_{\rm ECO}\sim
    \begin{cases} 
       e^{i\omega r_{*}} & \text{for}\ \quad r^*\to \infty 
       \\
       e^{-i\tilde{\omega} r_{*}}+\mathcal{R}_{\rm wall}e^{i\tilde{\omega} r_{*}} & \text{for} \quad r_{*}\to r_{\rm *(wall)}~.
    \end{cases}
\end{align}
Here, $\tilde{\omega}$ is the horizon-frame frequency as already defined and $r_{\rm *(wall)}$ denotes the location of the surface of the ECO in terms of the tortoise coordinate.  The reflectivity of the surface of the ECO is given by $\mathcal{R}_{\rm wall}$, which in general can be a function of the frequency $\omega$ as well. Using \ref{detweiler_fn}, which connects the Teukolsky radial perturbation variable ${}_{s}R_{\ell m}$ with ${}_{s}X_{\ell m}$, the $\mathcal{R}_{\rm wall}$ can also be related to $\mathcal{R}_{s}$. This is achieved by explicitly computing the ingoing and outgoing flux in the near horizon regime (at $r_{*}=r_{\rm *(wall)}$) using the Detweiler function and the Teukolsky radial perturbation variable. This yields \cite{Nakano:2017fvh,Wang_2018,Conklin:2019fcs}, 
\begin{align}\label{wall_ref}
|\mathcal{R}_{\rm wall}|^2=\frac{F^{\rm out}[r_{*}\to r_{\rm *(wall)}]}{F^{\rm in}[r_{*}\to r_{\rm *(wall)}]}
=
\begin{cases}
{|\mathcal{R}_{0}|^{2}\over |\mathcal{T}_{0}|^{2}} \qquad &s=0
\\
{|\mathcal{R}_{\pm1}|^{2}\over |\mathcal{T}_{\pm 1}|^{2}}\left|{\bar B^2\over E}\right|^{\pm1} \qquad &s=\pm 1
\\
{|\mathcal{R}_{\pm 2}|^{2}\over |\mathcal{T}_{\pm2}|^{2}}\left[{|C|^2 \over D}\right]^{\pm1}\qquad &s=\pm 2
\end{cases}
. 
\end{align}
where $|C|^{2}$ is related to the Starobinsky-Churilov constant \cite{Starobinsky:1974} and $\bar B^2$, $E$, and $D$ are dependent on $\omega$, $m$, the separation constant $\lambda$, as well as hairs of the black hole \cite{Nakano:2017fvh, Wang_2018}. Note that for the case of scalar perturbation neither $|C|^2$ nor $D$ has any effect and thus with $\mathcal{T}_{0}=1$, the reflectivity arising out of the Detweiler function is identical to that of the Teukolsky radial perturbation. On the other hand, for electromagnetic and gravitational perturbations, these terms will affect the reflectivity. Since we are not interested in an exact expression for the reflectivity but rather interested in understanding how a non-zero reflectivity can affect various physical characteristics of the solution, the exact relation in \ref{wall_ref} will not be of much importance. However, if some exact expression of the reflectivity can be obtained from some model of quantum gravity, then \ref{wall_ref} will of significant interest, as it can provide a direct connection between the macroscopic observations with the microscopic theoretical details. Since such is not the case, there is not much significant difference whether we call $\mathcal{R}_{\rm wall}$ or the ratio $(\mathcal{R}_{s}/\mathcal{T}_{s})$ as the reflectivity. For completeness, we will present all our results in terms of $\mathcal{R}_{\rm wall}$ using \ref{wall_ref}. 

For the reflectivity of the membrane (or of the quantum corrected apparent horizon) various choices can be made, namely --- (a) perfectly reflecting surface \cite{Maggio:2018ivz}, (b) partially reflecting surface with constant reflectivity and finally, (c) partially reflecting surface with a frequency dependent reflectivity $\mathcal{R}_{\rm wall}(\omega)$, a special case of which is the Boltzmann reflectivity \cite{Oshita:2019sat}. Among these models, the perfectly reflecting scenario is not very practical since they lead to ergoregion instability for moderate rotation of the ECO and also there are some observational constraints from the stochastic gravitational wave background \cite{Barausse:2018vdb}. Existence of the instability of modes for perfectly reflecting membranes is a generic one and exist in the present context as well (though much more tamed in the braneworld scenario than the Kerr-like ECOs, as we will see). Hence it is more favourable to work with models of the membrane with partial or frequency dependent reflectivity as described above. The reflectivity $\mathcal{R}_{\rm wall}$ in the constant reflectivity model of the membrane can be expressed as,
\begin{align}
\mathcal{R}_{\rm wall}=R_{\rm c}e^{i\delta_{\rm wall}}~;  \qquad 0<R_{\rm c}<1
\end{align}
where $\delta_{\rm wall}$ is a phase factor that depends on the model of the quantum gravity. The situation with $R_{\rm c}=0$ correspond to a black hole, while the case $R_{\rm c}=1$ is the perfectly reflecting membrane, which leads to instability and hence will not be considered here. On the other hand, if one assumes the black hole as a macroscopic realization of a discrete quantum system, then without going into  details of the quantum gravity theory the reflectivity of the membrane can be defined in terms of the Boltzmann factor as \cite{Oshita:2019sat,Wang:2019rcf},
\begin{align} \label{ref_b}
\mathcal{R}_{\rm wall}=e^{-|\tilde{\omega}|/(2T_{+})}e^{i\delta_{\rm wall}}~,
\end{align}
where $T_{+}$ is the Hawking temperature as defined in \ref{hawking_temp}. 

\begin{figure}[h]
\begin{center}
\includegraphics[width=10cm]{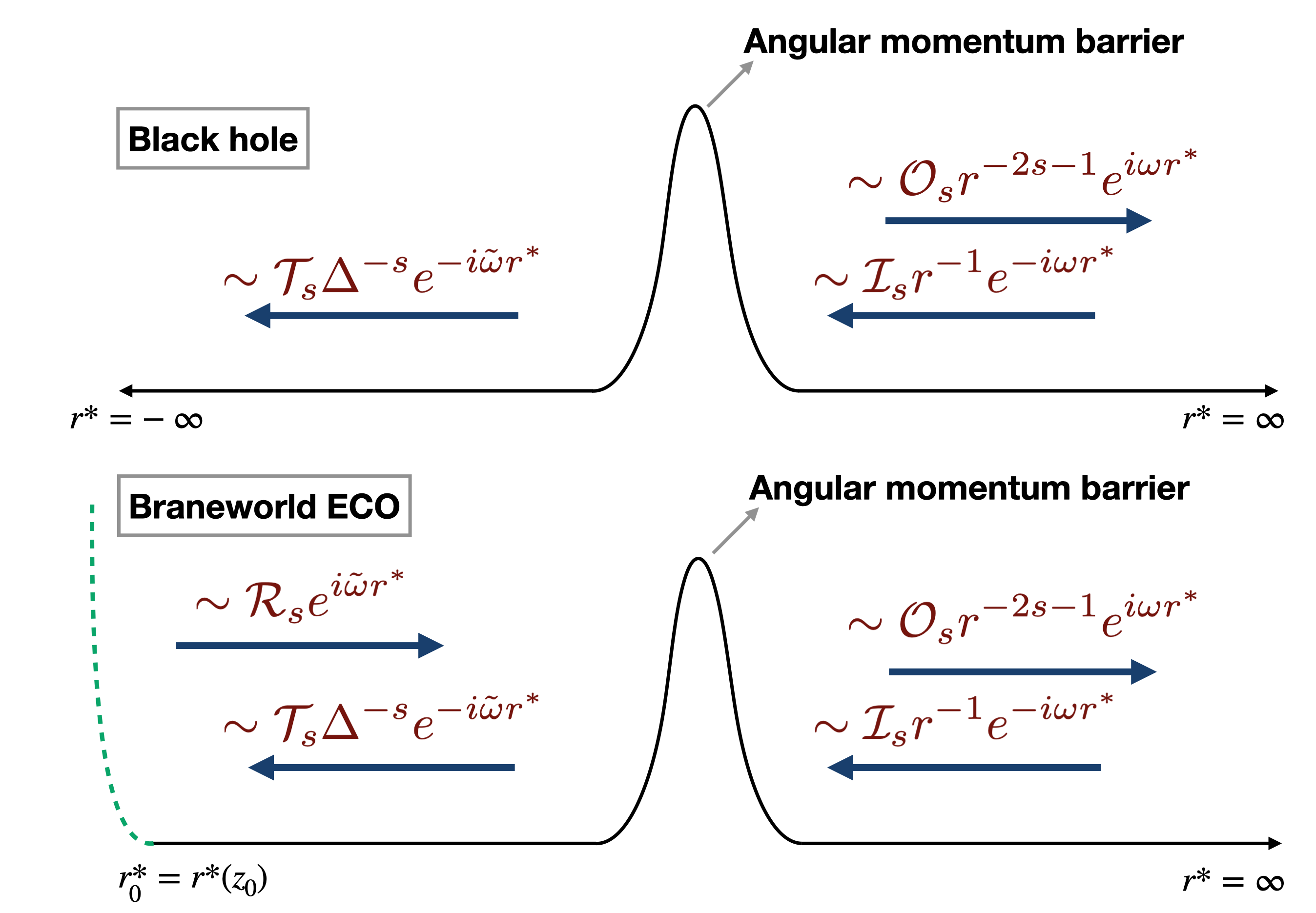}
\caption[]{The above figure presents the relevant boundary condition associated with the phenomenon of superradiance in the case of a black hole as well as a braneworld ECO. In the case of a black hole, the ingoing wave gets scattered by the angular momentum barrier and  goes to infinity while a part of it is transmitted. The scattered wave carries out more energy if the frequency of the mode satisfies the condition $\omega<m \Omega_{+}$. On the other hand, for braneworld ECO, there are both ingoing and outgoing modes in the near horizon regime, leading to possible runaway amplification of the amplitude of the ingoing wave from infinity, referred to as the ergoregion instability.}\label{figure:asymp}
 \end{center}
 \end{figure}

Finally, for superradiance there is an ingoing wave from infinity, a part of which gets reflected from the effective potential of the exterior geometry and goes back to infinity as an outgoing wave. A part of the incident wave is transmitted to the region inside the photon sphere (which is the location of the maxima of the effective potential) and ultimately hits the surface of the ECO. Then a part of it again gets reflected and hits the effective potential. From which a part goes to infinity while another part comes in and the whole  process is repeated. Therefore the boundary condition will involve both ingoing and outgoing waves at infinity as well as near the surface of the ECO, which suggests that it is given by \ref{asymp_ampli} and as shown in \ref{figure:asymp}. As we will show later, the amplification due to superradiance will depend on the reflectivity of the effective potential (which is identical to that of the black hole) and also on the reflectivity of the membrane/horizon. Thus the phenomenon of superradiance will also receive subtle modifications from the rotating ECO in the braneworld and will possibly make it more stable under superradiant instability due to the tidal charge.   

\section{Superradiant instability for exotic compact objects in braneworld}\label{superradiance}

For a rotating black hole, superradiance is the process by which the amplitude of an ingoing wave from infinity, falling on the black hole, is enhanced at the cost of the rotational energy of the black hole and is finally reflected outwards. For black holes this process does not lead to any instability (as the ingoing mode is always absorbed) while for ECOs, as we describe below, there is a potential instability. As the ingoing wave is incident on the black hole spacetime, it gets scattered by the effective potential. A part of it is reflected and goes to infinity, carrying the excess energy, while a part is transmitted and goes into the horizon. For ECOs, the part which is transmitted from the effective potential is further reflected by the surface of the ECO and since it passes through the ergoregion, it can extract rotational energy from the ECO. A part of this wave is again reflected and goes to the surface of the ECO with an increased amplitude. Then the whole process repeats, leading to a runaway situation, where the amplitude of the modes inside the ergoregion grows indefinitely. Thus for a spinning ECO, with a reflective membrane in front of the horizon, the process of superradiance may lead to a potential instability. This is known as the superradiant instability or ergoregion instability \cite{Brito:2015oca}. For clarity, we have provided a pictorial depiction of the physical scenario discussed above in \ref{figure:asymp}. 

To reiterate, by definition negative energy states can exist within the ergosphere and even though it is energetically more favourable to drift towards even more negative energy states, the black holes are stable. This is because, the presence of an event horizon makes the black hole absorb all the ingoing radiation very effectively. However, in the case of an ECO due to the existence of a reflective surface such dissipation mechanism is absent, making them highly unstable. The instability is enhanced for perturbing fields with higher spin and hence for gravitational perturbations we except the problem of superradiant instability to be the strongest. As we will see, presence of a non-zero tidal charge curbs the instability.  

\begin{figure}[!ht] 
\begin{center}
\includegraphics[scale=0.25]{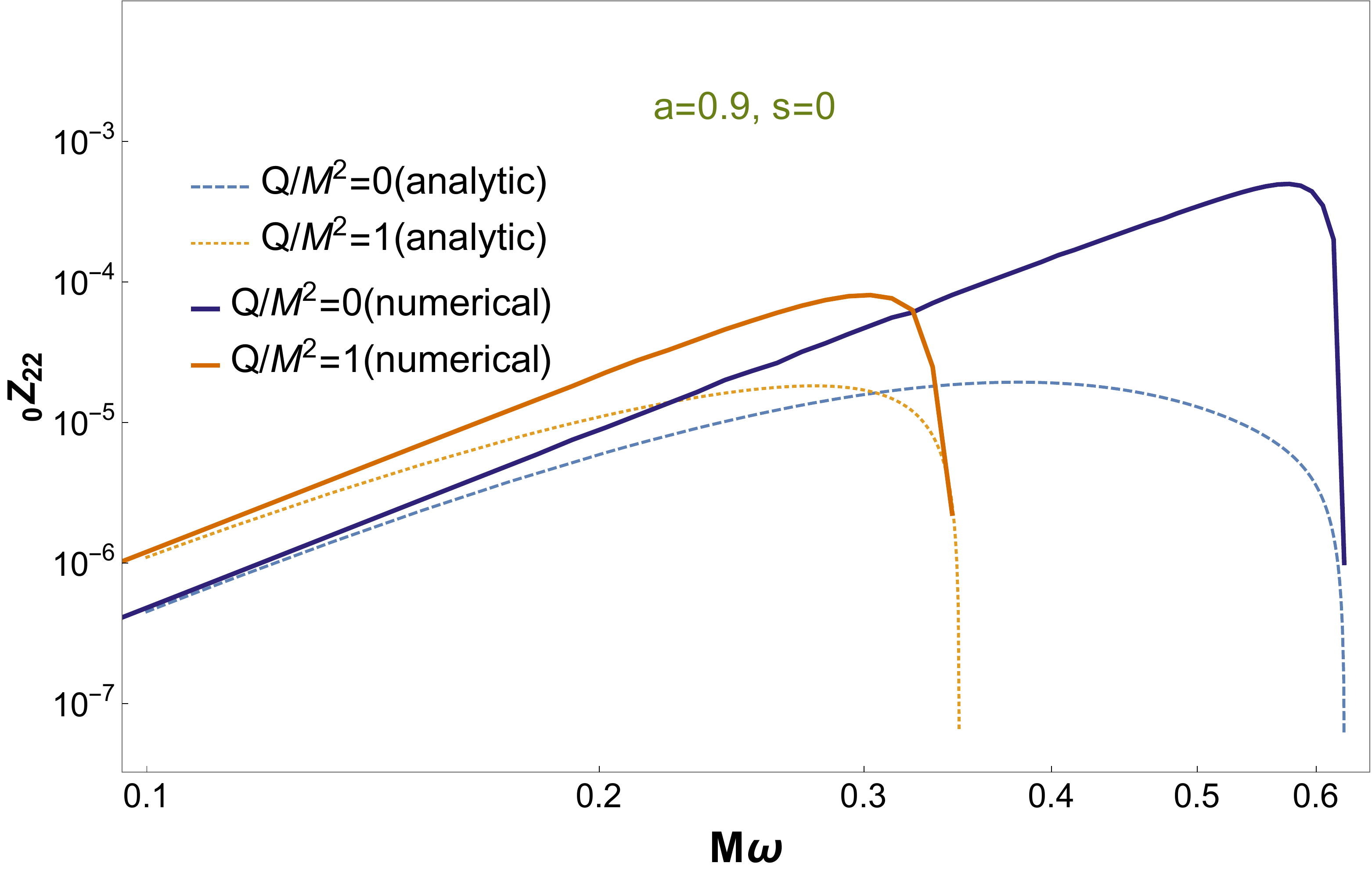}~~
\includegraphics[scale=0.25]{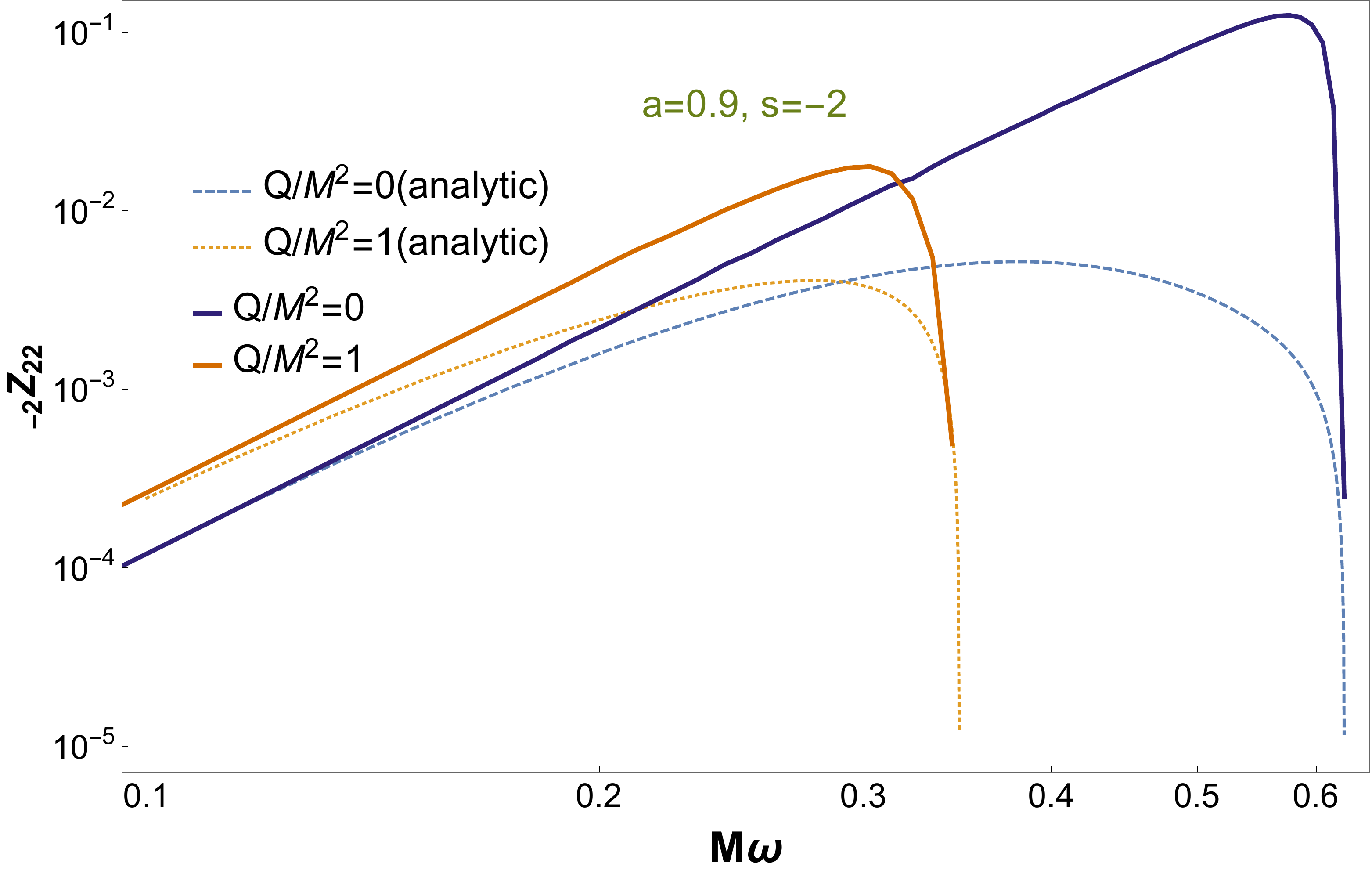}
\caption{The amplification factor ${}_{s}Z_{\ell m}$ is plotted against the frequency both from the analytical expression and numerical estimation, for different values of the dimensionless tidal charge parameter $(Q/M^{2})$. The plot of the amplification factor ${}_{s}Z_{\ell m}$ for the scalar case ($s=0$) is plotted to the left and for the gravitational case ($s=-2$) is plotted to the right. We see from the above plots that with an increase in the tidal charge parameter $Q$, the ergoregion instability is suppressed. This is because, the critical frequency upto which the amplification happens decreases, as well as the maximum amplification factor also decreases drastically. Though at a given frequency, the amplification in the presence of extra dimension is slightly larger compared to the case of Kerr black hole. See text for more discussions.}
\label{figure_super}
\end{center}
\end{figure}

The best way to quantify the phenomenon of superradiance is through the amplification of the ingoing wave from infinity. As evident from \ref{asymp_ampli}, the amplification factor ${}_{s}Z_{\ell m}$ will depend on the ratio $|\mathcal{O}_{s}/\mathcal{I}_{s}|$ and it can be defined in terms of the ratio of ingoing and outgoing energy fluxes at infinity as \cite{Brito:2015oca},
\begin{align}\label{ampli_super}
{}_{s}Z_{\ell m}={dE_{\rm out}\over dE_{\rm in}}-1
=
\begin{cases}
\left|{\mathcal{O}_{0}\over \mathcal{I}_{0}}\right|^2-1,\qquad &s=0
\\
\left|{\mathcal{O}_{1}\over \mathcal{I}_{1}}\right|^2\left({16\omega^4\over \bar B^2}\right)^{\pm 1}-1,\qquad &s=\pm1
\\
\left|{\mathcal{O}_{2}\over \mathcal{I}_{2}}\right|^2\left({256\omega^8\over |C|^2}\right)^{\pm 1}-1,\qquad &s=\pm2
\end{cases}
\end{align}
where, $|C|$ is related to the Starobinsky-Churilov constant and $|\bar B|$ is dependent on $\omega$, $m$, the separation constant $\lambda$ and the asymptotic hairs of the ECOs. Based on the transformation of the radial perturbation variable of Teukolsky to Detweiler function, the amplification factor ${}_{s}Z_{\ell m}$ can be written in terms of the Detweiler asymptotic amplitudes as
\begin{align}\label{super_boundary}
{}_{s}Z_{\ell m}=|\mathcal{R}_{\rm BH}|^{2}-1~;\qquad {}_{s}X_{\ell m}(r_{*}\rightarrow \infty )=\mathcal{R}_{\rm BH}e^{i\omega r_{*}}+e^{-i\omega r_{*}}~,
\end{align}
As we have emphasized earlier, it is difficult to solve for the radial perturbation equation analytically in some generic context  and the same consideration applies to the present context as well. It is possible to determine the amplification factor ${}_{s}Z_{\ell m}$ analytically, in the low frequency regime \cite{Starobinsky:1973aij, Starobinsky:1974,Page:1976df, Brito:2015oca}. For the background spacetime, given by the rotating solution on the brane with a non-zero tidal charge, the amplification factor for a generic spin-$s$ field can be given by (for a detailed derivation, see \ref{AppSuper}),
\begin{align} \label{Z_analytic}
{}_{s}Z_{\ell m}=|\mathcal{R}_{\rm BH}|^{2}-1
=4\sigma\left[{(\ell-s)!(\ell+s)!\over (2\ell)!(2\ell+1)!!}\right]^{2}\left[\omega(r_+-r_-)\right]^{2\ell+1}\prod_{n=1}^{\ell}\bigg(1+{4\sigma^{2}\over n^{2}}\bigg),
\end{align}
where, the quantity $\sigma$ takes the following form, $\sigma=\{(r_{+}^{2}+a^{2})/(r_{+}-r_{-})\}(m\Omega_{+}-\omega)$, which explicitly depends on the tidal charge parameter $Q$. We can see from the above expression for the amplification factor ${}_{s}Z_{\ell m}$, that for $\omega<m\Omega_{+}$, the amplification factor is positive, i.e., the amplitude of the outgoing mode is larger than the ingoing mode, i.e., within the superradiant bound (this corresponds to $\omega<m\Omega_{+}$) the reflectivity of the black hole is greater than unity. Using the above expression for $\sigma$, we can write down the above amplification factor explicitly in the low frequency regime as, 
\begin{align} \label{Z_analyticQ}
{}_{s}Z_{\ell m}&=4M\left(\frac{1+\frac{Q}{2M^{2}}+\sqrt{1-\chi^{2}+\frac{Q}{M^{2}}}}{\sqrt{1-\chi^{2}+\frac{Q}{M^{2}}}}\right)\left[\frac{m}{2M}\left(\frac{\chi}{1+\frac{Q}{2M^{2}}+\sqrt{1-\chi^{2}+\frac{Q}{M^{2}}}}\right)-\omega\right]
\nonumber
\\
&\hskip 0.5 cm \times \left[{(\ell-s)!(\ell+s)!\over (2\ell)!(2\ell+1)!!}\right]^{2}\left[2M\omega\sqrt{1-\chi^{2}+\frac{Q}{M^{2}}}\right]^{2\ell+1}
\nonumber
\\
&\hskip 1 cm \times \prod_{n=1}^{\ell}\Bigg\{1+\frac{4M^{2}}{n^{2}}\left(\frac{1+\frac{Q}{2M^{2}}+\sqrt{1-\chi^{2}+\frac{Q}{M^{2}}}}{\sqrt{1-\chi^{2}+\frac{Q}{M^{2}}}}\right)^{2}\left[\omega-\frac{m}{2M}\left(\frac{\chi}{1+\frac{Q}{2M^{2}}+\sqrt{1-\chi^{2}+\frac{Q}{M^{2}}}}\right)\right]^{2}\Bigg\}
\end{align}
The above analysis provides the desired analytical expression for the amplification factor due to superradiance. Note that the amplification factor depends explicitly on the tidal charge parameter $Q$ and reduces to that of a Kerr black hole in the limit of vanishing tidal charge. In generic contexts, one has to resort to numerical computation in order to determine the amplification factor and \ref{ampli_super} plays a key role in that. In particular, using a modified version of the Mathematica code used in \cite{Brito:2015oca}, we have numerically determined the amplification factor by computing the asymptotic amplitudes $\mathcal{O}_{s}$ and $\mathcal{I}_{s}$, respectively. Hence the amplification factor has been determined in a case by case basis using \ref{ampli_super}. In order to see the consistency of the analytical result derived in \ref{Z_analyticQ} with the numerical analysis, the amplification factor arising out of numerical as well as analytical treatment of superradiance has been plotted as a function of the frequency $\omega$ in \ref{figure_super}. As evident, ${}_{s}Z_{\ell m}>0$ for the superradiant frequencies, irrespective of the spin $s$ of the perturbation. In addition, \ref{figure_super} explicitly demonstrates the usefulness of the tidal charge parameter $Q$ in taming the superradiant instability. As the value of the tidal charge parameter is increased, the critical frequency $m\Omega_{+}$ decreases in comparison to the scenario with zero tidal charge. Hence the parameter space allowed in the frequency domain decreases by almost $45\%$, for $(Q/M^{2})=1$, in comparison to the case of vanishing tidal charge, see \ref{figure_super}. This is because, the presence of the extra dimension and its back-reaction on the brane makes the gravitational pull stronger, thereby increasing the size of the would-be-horizon and decreasing the angular velocity $\Omega_{+}$, which in turn lowers the critical frequency $m\Omega_{+}$. Besides reducing the parameter space in the frequency domain, presence of extra dimension also helps in reducing the maxima of the amplification factor. For gravitational perturbation, the maximum amplification factor in the presence of extra dimension, with $(Q/M^{2})=1$, is about ten times smaller to the case with $(Q/M^{2})=0$, as \ref{figure_super} explicitly demonstrates. Even though, for a given frequency $\omega$, the amplification in the presence of extra dimension is larger in comparison to the general relativistic scenario, the reduction in the parameter space in the frequency domain and decrease in the maximum amplification by a significant amount makes the rotating ECO more stable against the superradiant instability. As we will see, an identical scenario holds true for the static modes as well, to be discussed in the later sections.

For a horizonless ECO, as pointed out before, if the horizon is replaced by a perfectly reflective membrane, whenever the amplification factor ${}_{s}Z_{\ell m}>0$, it would give rise to superradiant instability. The `bounce and amplify' picture for the above phenomenon has also been discussed before. To see how this assertion comes about, note that the ingoing wave after transmission at the effective potential barrier gets reflected by the surface of the ECO and hence its amplitude will be modified by a factor of $\mathcal{R}_{\rm wall}$. Subsequently, it is again reflected by the potential barrier and then it will hit the surface of the ECO with amplitude $\mathcal{R}_{\rm wall}\mathcal{R}_{\rm BH}$. After which this process will repeat. Therefore, the wave gets amplified by a factor of $|\mathcal{R}_{\rm wall}\mathcal{R}_{\rm BH}|$ due to each bounce at the surface of the ECO. Thus, the energy of the perturbation mode would grow indefinitely unless $|\mathcal{R}_{\rm wall}\mathcal{R}_{\rm BH}|<1$.  If we consider a perfectly reflecting surface, with $\mathcal{R}_{\rm wall}=1$, then the above condition can never be satisfied. As shown in \cite{Maggio:2018ivz}, a way to curb this instability is to allow some absorption of the ingoing radiation by the surface of the ECO, i.e., we would like to choose $\mathcal{R}_{\rm wall}<1$, such that the condition $|\mathcal{R}_{\rm wall}\mathcal{R}_{\rm BH}|<1$ can be satisfied. This will allow us to put possible constraints on the reflectivity of the membrane, in terms of the amplification due to superradiance, which yields,
\begin{align} \label{max_rwall}
|\mathcal{R}_{\rm wall}|^2<{1\over 1+{}_{s}Z_{\ell m}}.
\end{align}
This bound on the reflectivity of the surface of the ECO depends crucially on the tidal charge parameter $Q$ through \ref{Z_analyticQ}. As emphasized before, this ensures that the energy inside the cavity formed by the angular momentum barrier and the surface of the ECO does not grow indefinitely. Based on the numerical computation of the amplification factor, we provide the maximum allowed values for the reflectivity $\mathcal{R}_{\rm wall}$ of the surface of the ECO,  in \ref{ref_bound}, so that the superradiant instability can be quenched. 

\begin{table}
\centering
\begin{tabular}{ |c||c|c| } 
\hline
\multicolumn{3}{|c|}{Bounds on $|\mathcal{R}_{\rm wall}|$} \\
\hline
$a$ & $(Q/M^{2})=0$& $(Q/M^{2})=1$ \\
\hline
0.998 & 0.64  & 0.72
\\ 
0.9 & 0.94 & 0.98
\\ 
0.8& 0.992 &0.999
\\ 
\hline
\end{tabular}
\caption{We have presented various bounds on the reflectivity $\mathcal{R}_{\rm wall}$, of the surface of the ECO, for different values of the black hole spin parameter $a$ and the tidal charge parameter $Q$, obtained from \ref{max_rwall}, for the following choices of the parameters: $s=-2$, $\ell=2$ and $m=2$ respectively, based on the numerical computation of ${}_{s}Z_{\ell m}$. As evident, for a given rotation parameter, the reflectivity increases with an increase of the tidal charge parameter $Q$, i.e., as the size of the extra dimension decreases.}
\label{ref_bound}
\end{table}

\begin{figure}
\begin{centering}
\includegraphics[scale=0.25]{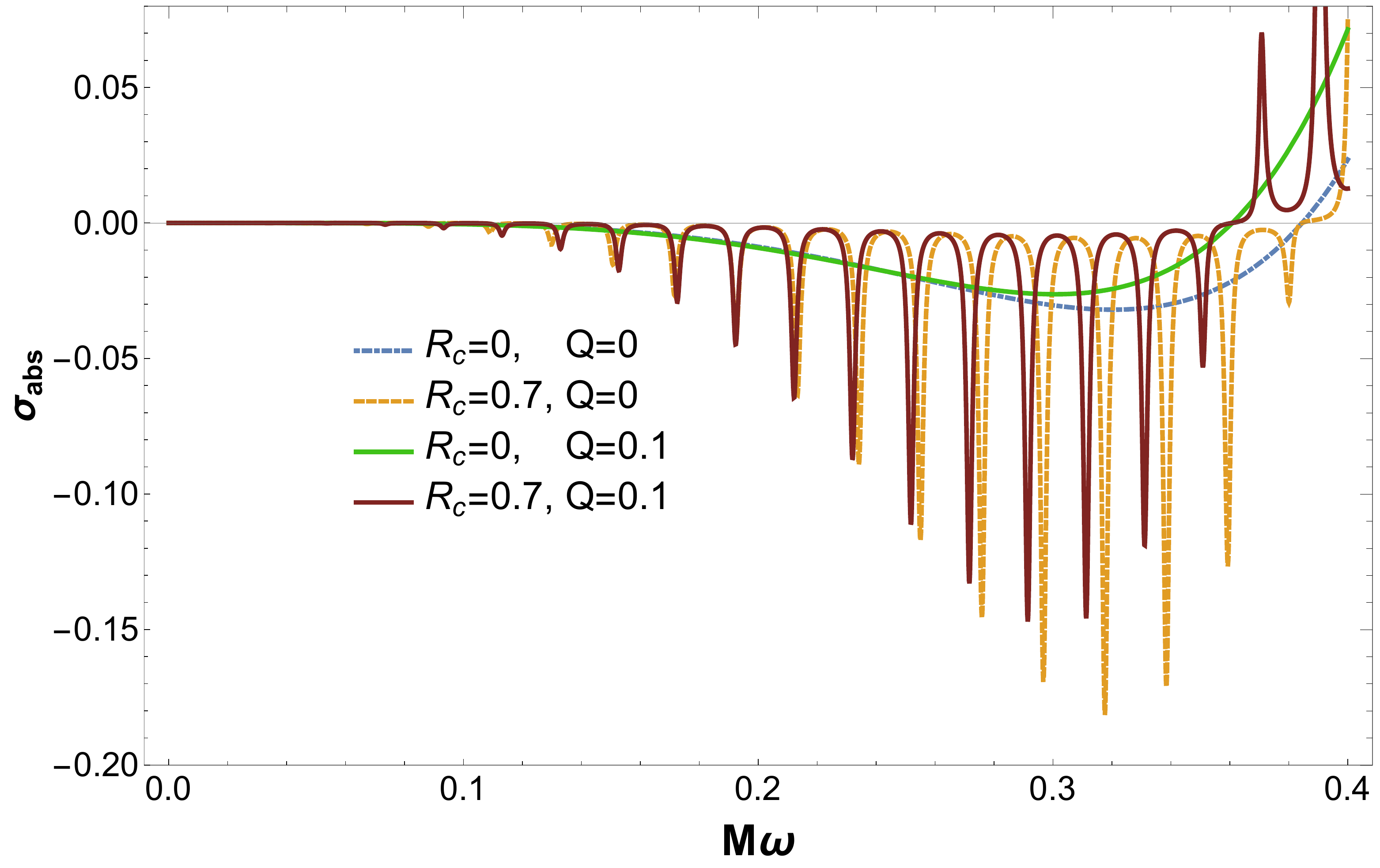}
\includegraphics[scale=0.25]{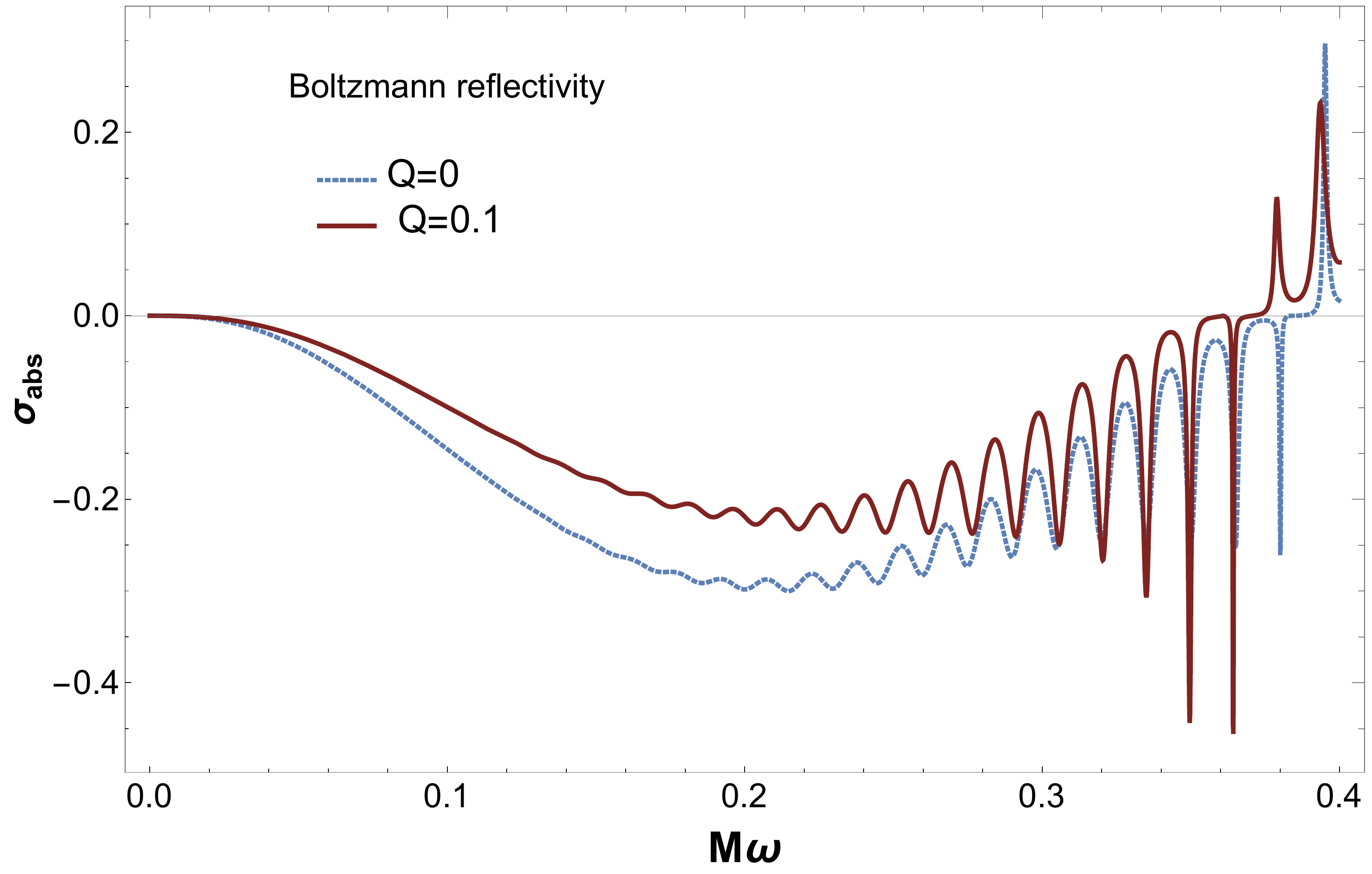}
\caption[]{The absorption cross-section $\sigma_{\rm abs}$ of the surface of the ECO is plotted against the frequency of the perturbation modes for different values of the reflectivity and the tidal charge. The zero crossing of the absorption cross-section corresponds to the superradiance bound $\omega=m\Omega_{+}$. See text for more discussions.}\label{figure:Z}
\end{centering}
\end{figure}

As the phenomenon of superradiance and the preceding discussion explicitly demonstrates that in order to avoid ergoregion instability it is necessary for the surface of the ECO to absorb some amount of ingoing radiation. This can be quantified in terms of the absorption cross-section, defined as $\sigma_{\rm abs}$, which is defined as,
\begin{equation}
\sigma_{\rm abs}={F^{\rm in}[r_{*}\to r_{\rm *(wall)}]+F^{\rm out}[r_{*}\to r_{\rm *(wall)}]\over F^{\rm in}[r_{*}\to \infty]}~.
\end{equation}
The absorption cross section $\sigma_{\rm abs}$ can be computed analytically in the low frequency approximation, using the solutions of the radial perturbation equation in this limit, as shown is \ref{abs_app}. However, it is more useful to perform a numerical analysis to depict the effect of the reflectivity of the surface of the ECO and the tidal charge on the absorption cross section, which has been presented in \ref{figure:Z}. As evident, presence of the tidal charge increases the absorption cross-section, thereby reducing the ergoregion instability compared to the general relativistic scenario. Thus the absorption cross-section also provides us a similar conclusion, i.e., the presence of the tidal charge, or, equivalently that of an extra spatial direction, reduces the ergoregion instability for exotic compact objects living on the brane. 

\section{Static modes and Darboux transformation for exotic compact object on the brane}\label{static_mode_ECO}

In relation to the ergoregion instability, we have seen that the low frequency modes play a crucial role and in particular the $\omega=0$, i.e., the static mode is most important in this respect. In addition, the instability is enhanced if the surface of the ECO is perfectly reflecting. Following which, in this section we discuss the static perturbation modes for generic spin for the braneworld ECO with perfectly reflecting surface. In particular, we will see that the critical value of the rotation parameter depends on the tidal charge parameter. To start with, we rewrite \ref{radial_perturbation} for the static modes as,
\begin{align}
\Delta^{-s}\frac{d}{dr}\bigg(\Delta^{s+1}\dfrac{d{}_{s}R_{\ell m}}{dr}\bigg)&+{1\over \Delta}\bigg[a^{2}m^{2}+isam\Delta'-\Delta\lambda \bigg] {}_{s}R_{\ell m}=0~,
\label{radial_perturbation_static}
\end{align}
where, $\lambda=(\ell-s)(\ell+s+1)$ follows from the angular perturbation equation. As we will show later, the solution of the above equation for gravitational and electromagnetic perturbation can be related to the scalar perturbation through the Darboux transformation. Thus we will discuss the solution of the above equation for scalar perturbation first, and then shall comment on the electromagnetic and gravitational perturbation using Darboux transformation. 
\subsection{Static mode for scalar perturbation}

For the scalar perturbation, with $s=0$, the radial perturbation equation for the static mode can be expressed as,
\begin{align}
\frac{d}{dr}\bigg(\Delta\dfrac{d{}_{0}R_{\ell m}}{dr}\bigg)&+{1\over \Delta}\bigg[a^{2}m^{2}-\ell(\ell+1)\Delta \bigg] {}_{0}R_{\ell m}=0~,
\label{perturbation_static_scalar}
\end{align}
Introducing, $z\equiv (r-r_{+})/(r_{+}-r_{-})$ and expressing $\Delta=(r-r_{+})(r-r_{-})=z(z+1)(r_{+}-r_{-})^{2}$ the above differential equation can be written as,
\begin{align}
\frac{d}{dx}\bigg[\left(1-x^{2}\right)\dfrac{d{}_{0}R_{\ell m}}{dx}\bigg]&+\bigg[\ell(\ell+1)-\frac{\left(i\nu\right)^{2}}{1-x^{2}} \bigg] {}_{0}R_{\ell m}=0~,
\end{align}
where, we have introduced a new variable $x=1+2z$ and a new constant, $\nu\equiv 2am/(r_{+}-r_{-})$. The above differential equation is identical to the Legendre differential equation and hence the general solution is in terms of the associated Legendre polynomials $P_{\ell}^{i\nu}(x)$ and $Q_{\ell}^{i\nu}(x)$. However, regularity of the perturbation as $r\rightarrow \infty$ demands the coefficient of $P_{\ell}^{i\nu}(x)$ to vanish. Given the perfectly reflective nature of the surface of the ECO, imposing Dirichlet boundary condition at the surface of the ECO leads to $Q_{\ell}^{i\nu}(1+2z_{0})=0$, while the Neumann boundary condition yields, $(d/dz)Q_{\ell}^{i\nu}(1+2z)\vert_{z_{0}}=0$. This yields,
\begin{align}\label{critical_a}
z_{0}^{-i\nu}\approx e^{i(p+1)\pi}\frac{\Gamma(1-i\nu)\Gamma(\ell+1+i \nu)}{\Gamma(1+i\nu)\Gamma(\ell+1-i \nu)}~,
\end{align}
where $p$ is an odd integer for the Dirichlet boundary condition and $p$ is an even integer for the Neumann boundary condition. The above algebraic relation determines the critical value of the rotation parameter $a=a_{\rm crit}$, beyond which the ECO will experience superradiant instability. Analytical estimate of $a_{\rm crit}$ for braneworld ECO can be obtained from \ref{critical_a} using the $\nu\rightarrow 0$ limit, which is equivalent to slow rotation limit. This yields,
\begin{align}\label{crit_a_stat}
a_{\rm crit}=\frac{\pi\left(p+1\right)M}{m |\ln \epsilon |}\sqrt{1+\frac{Q}{M^{2}}}~.
\end{align}
Note that the dimensionless parameter $\epsilon$, appearing in the above expression, denotes the departure of the surface of the ECO from the horizon location of the would-be black hole, which for the case of braneworld ECO is given by \ref{rp_shift}. As the above expression suggests, presence of a positive tidal charge $Q$, whose origin is solely from the existence of an extra spatial dimension, indeed suppresses the instability by increasing the value of $a_{\rm crit}$, as evident from \ref{crit_a_stat}. This is consistent with our earlier findings, where the presence of a tidal charge reduces the amplification due to superradiance. Thus we can safely conclude that braneworld ECOs are more stable than other candidates for ECO, as far as static modes are concerned. We will now present the results for electromagnetic and gravitational perturbation in the context of static modes. 

\subsection{Darboux transformation and static mode for electromagnetic and gravitational perturbation}

For static modes the electromagnetic and gravitational perturbations are closely related to the scalar perturbation and its derivative through the Darboux transformation. For the electromagnetic perturbation, the radial perturbation equation for the static modes, presented in \ref{radial_perturbation_static} takes the following form,
\begin{align}
\Delta\frac{d}{dr}\bigg(\dfrac{d{}_{-1}R_{\ell m}}{dr}\bigg)&+{1\over \Delta}\bigg[-\ell(\ell+1)\Delta+a^{2}m^{2}-iam\Delta'\bigg] {}_{-1}R_{\ell m}=0~,
\label{static_elec}
\end{align}
which can be related to the scalar perturbation equation for the static mode, through the following transformation,
\begin{align}\label{elec_sca}
{}_{-1}R_{\ell m}={}_{0}R_{\ell m}+\frac{i\Delta}{am}\dfrac{d{}_{0}R_{\ell m}}{dr}~,
\end{align}
as one can check in a straightforward manner. Therefore, the calculation for the scalar perturbation will be directly applicable for the case of electromagnetic perturbation as well. In an identical manner, the radial perturbation equation for the gravitational perturbation, from \ref{radial_perturbation_static}, can be expressed as,
\begin{align}\label{grav_stat}
\Delta^{2}\frac{d}{dr}\bigg(\frac{1}{\Delta}\dfrac{d{}_{-2}R_{\ell m}}{dr}\bigg)&+{1\over \Delta}\bigg[a^{2}m^{2}-2iam\Delta'-(\ell-1)(\ell+2)\Delta\bigg] {}_{-2}R_{\ell m}=0~.
\end{align}
The gravitational perturbation ${}_{-2}R_{\ell m}$ can be related to the electromagnetic perturbation through the following relation,
\begin{align}\label{grav_elec}
{}_{-2}R_{\ell m}=\frac{\left(am-i\Delta'\right)}{2am-i(r_{+}-r_{-})}{}_{-1}R_{\ell m}+\frac{i\Delta}{2am-i(r_{+}-r_{-})}\dfrac{d{}_{-1}R_{\ell m}}{dr}~,
\end{align}
since, it is straightforward to check that ${}_{-2}R_{\ell m}$ indeed satisfies \ref{grav_stat}, provided ${}_{-1}R_{\ell m}$ satisfies \ref{static_elec}. Note that these relations are not unique, as there can be arbitrary overall multiplicative factors. Given these relations, we can immediately determine the appropriate boundary conditions for the electromagnetic and gravitational perturbations, using the boundary condition on the scalar perturbation. These can be obtained by inverting the relations presented in \ref{elec_sca} and \ref{grav_elec} and expressing ${}_{0}R_{\ell m}$ and its derivative in terms of ${}_{-1}R_{\ell m}$ and ${}_{-2}R_{\ell m}$, along with their derivatives. These relations, when expressed in an explicit manner, yields,
\begin{align}
{}_{0}R_{\ell m}&=-\frac{iam}{\ell(\ell+1)}\Bigg[\left(\frac{iam}{\Delta}\right){}_{-1}R_{\ell m}+\dfrac{d{}_{-1}R_{\ell m}}{dr}\Bigg]~,
\label{cond_01}
\\
\Delta \dfrac{d\,{}_{0}R_{\ell m}}{dr}&=-\frac{iam}{\ell(\ell+1)}\Bigg[iam \dfrac{d{}_{-1}R_{\ell m}}{dr}+{}_{-1}R_{\ell m}\left(\frac{\ell(\ell+1)\Delta-a^{2}m^{2}}{\Delta} \right)\Bigg]~,
\label{cond_02}
\end{align}
for electromagnetic perturbation. Therefore, the Dirichlet boundary condition demands setting ${}_{0}R_{\ell m}=0$ on the surface of the ECO and hence the combination within square bracket of \ref{cond_01} must vanish. Similarly, for the Neumann boundary condition, we can set $(d\,{}_{0}R_{\ell m}/dr)=0$, and hence the electromagnetic perturbation ${}_{-1}R_{\ell m}$ must satisfy the differential relation, arising out of vanishing of the terms within the square bracket of \ref{cond_02}. Furthermore, we can use these boundary conditions on the electromagnetic perturbation in order to determine the relevant boundary condition on the gravitational perturbation using \ref{grav_elec}. These yield, 
\begin{align}
\dfrac{d{}_{-2}R_{\ell m}}{dr}&+\Bigg[\frac{iam}{\Delta}+\frac{(\ell+2)(\ell-1)}{2iam+\Delta'} \Bigg]{}_{-2}R_{\ell m}=0~,
\quad \textrm{for~the~Dirichlet~boundary~condition}
\\
\dfrac{d{}_{-2}R_{\ell m}}{dr}&+\Bigg[\frac{iam}{\Delta}+\frac{iam(\ell-1)(\ell+2)}{iam(2iam+\Delta')+\ell(\ell+1)\Delta}\Bigg]{}_{-2}R_{\ell m}=0~,
\quad \textrm{for~the~Neumann~boundary~condition}
\end{align}
as the boundary conditions for the gravitational perturbation on the surface of the ECO. Therefore, the computation of the critical rotation parameter $a_{\rm crit}$ for the scalar perturbation can be easily extended to both electromagnetic and gravitational perturbations with appropriate boundary conditions, as presented above. As evident, for the static modes associated with both electromagnetic and gravitational perturbations, the boundary conditions for the ECO on the brane are affected by the presence of extra dimension, through the tidal charge parameter $Q$ appearing in $\Delta$. Also, the value of $a_{\rm crit}$ will increase in the presence of extra dimension, as it predicts a positive value of the tidal charge parameter $Q$, thereby stabilizing the ECO against the superradiant instability for electromagnetic and gravitational perturbations.

\section{Spectrum of the quasi-normal modes for the exotic compact object on the brane: Analytical results}\label{qnm_analytical}

In this section, we will determine the QNMs associated with the perturbation of the ECO using analytical techniques. This requires solving the radial and angular perturbation equations, presented in \ref{radial_perturbation} and \ref{angular_perturbation}, respectively. As emphasized earlier, for obtaining the QNM spectrum, we need to impose certain boundary conditions at the horizon and also at the asymptotic infinity, which has been discussed in detail in \ref{bound_QNMT} and \ref{echo_sol} (also see \ref{figure:asymp_qnm}). In brief, one must assume that there are no ingoing waves from the asymptotic infinity, which is unanimously true for both the black hole and the ECO. While, for a black hole the boundary condition at the horizon is ingoing, but as we have discussed, for the horizonless ECO the boundary condition at the horizon is modified due to the presence of the reflective surface in the near horizon region.

Using these boundary conditions in the low-frequency regime $(M\omega \ll 1)$, it is possible to obtain the characteristic frequencies by matching the solution in the near-horizon region with the solution near infinity. In the near-horizon region, described by the condition $\omega(r-r_{+})\ll 1$, the radial perturbation equation can be simplified  by the low-frequency approximation, leading to the following solution in terms of Hypergeometric functions (see \ref{app_perturbation}),
\begin{align} \label{sol_near}
{}_{s}R_{\ell m}&=(1+z)^{i\sigma}\Big(Az^{-i\sigma}~{}_{2}F_{1}[-\ell+s,\ell+1+s;1-2i\sigma+s;-z]
\nonumber
\\
&\hskip 3 cm +Bz^{i\sigma-s}~{}_{2}F_{1}[-\ell+2i\sigma,\ell+1+2i\sigma;1+2i\sigma-s;-z]\Big).
\end{align}
Here, $A$ and $B$ are unknown constants, which need to be determined using appropriate boundary conditions. We have also used the following definitions: $z\equiv (r-r_{+})/(r_{+}-r_{-})$ and $\sigma \equiv (r_{+}^{2}+a^{2})(m\Omega_{+}-\omega)/(r_{+}-r_{-})$. On the other hand, in the asymptotic region $r\gg M$, the radial perturbation equation can be reduced to a simple form in the low-frequency approximation, which can be immediately solved, yielding (see \ref{app_perturbation}),
\begin{align} \label{infty_soln}
{}_{s}R_{\ell m}=\alpha e^{-ikz}z^{\ell-s} U(\ell-s+1,2\ell+2,2ik z)+\beta e^{-ik z}z^{-\ell-s-1}U(-\ell-s,-2\ell,2ikz)
\end{align}
where, $\alpha$ and $\beta$ are two unknown constants to be determined using the boundary conditions and $k\equiv \omega(r_{+}-r_{-})$, with $z$ defined earlier. If we take the $r\rightarrow \infty$ limit of \ref{infty_soln}, we will observe that it will provide both ingoing as well as outgoing solutions. Since by the boundary condition, presented in \ref{bound_QNMT}, there will be no ingoing wave at infinity, we obtain the following condition among the arbitrary constants $\alpha$ and $\beta$ as, 
\begin{align}\label{far_coefficient}
\frac{\beta}{\alpha}=-{\Gamma(2\ell+2)\Gamma(-\ell+s)\over \Gamma(\ell+s+1)\Gamma(-2\ell)}\left(-2ik\right)^{-1-2\ell}.
\end{align}
Substitution of this condition in \ref{infty_soln} will eliminate one arbitrary constant, leaving an overall factor. Subsequently, one takes the small $(r-r_{+})$ limit of \ref{infty_soln} and match it with the $r\rightarrow \infty$ limit of \ref{sol_near}, in the intermediate overlapping region $(M\ll r-r_{+}\ll 1/\omega)$, yielding,
\begin{align} \label{A_by_B}
{A\over B}=-{\Gamma(\ell+1+s)\over \Gamma(\ell+1-s)}
\bigg[{R_{+}+i(-1)^{\ell}k^{2\ell+1}LS_{+}\over R_{-}+i(-1)^{\ell}k^{2\ell+1}LS_{-}}\bigg],
\end{align}
where we have defined the following,
\begin{align}
R_{\pm}={\Gamma(1\pm2i\sigma \mp s)\over \Gamma(\ell+1\pm 2i\sigma)},
\quad S_{\pm}={\Gamma(1\pm 2i\sigma \mp s)\over \Gamma(-\ell\pm 2i\sigma)}, 
\quad
L={1\over 2}\bigg[{2^{\ell}\Gamma(\ell+1+s)\Gamma(\ell+1-s)\over \Gamma(2\ell+1)\Gamma(2\ell+2)}\bigg]^2.
\end{align}
Finally, the determination of the QNMs requires imposing relevant boundary condition at the near-horizon regime. Conventionally for a black hole spacetime, an ingoing boundary condition is imposed and therefore one obtains another relation between $A$ and $B$, which along with \ref{A_by_B} determines the QNMs. As we have discussed for ECO, where the horizon is replaced by a partially reflective membrane in front of the would be horizon, the relevant boundary condition corresponds to \ref{bound_QNMT}. This when applied to the near region solution presented in \ref{sol_near}, yields the ratio $(\mathcal{R}_{s}/\mathcal{T}_{s})$ in terms of the ratio $(A/B)$ of the unknown constants appearing in the near region solution of the Teukolsky equation. Using this equation along with the ratio $(A/B)$ given in \ref{A_by_B}, the reflectivity $\mathcal{R}_{\rm wall}$ can be obtained from \ref{wall_ref}. Since the reflectivity $\mathcal{R}_{\rm wall}$ depends on the spin $s$ of the perturbation, we will discuss the QNM spectrum for braneworld ECO, individually for each spin.

\subsection{Quasi-normal mode frequencies for scalar perturbation}\label{qnm_scalar_sec}

The scalar perturbations are defined by $s=0$, for which the ratio $\mathcal{R}_{0}/\mathcal{T}_{0}$ is given by the ratio $(A/B)$, modulo some additional phase factor. Similarly, the reflectivity of the membrane $\mathcal{R}_{\rm wall}$, is related to the ratio $\mathcal{R}_{0}/\mathcal{T}_{0}$, with some arbitrary phase factor, as evident from \ref{wall_ref}. Finally the reflectivity of the surface of the ECO takes the following form (see \ref{app_perturbation} for the derivation of $(\mathcal{R}_{0}/\mathcal{T}_{0})$),
\begin{align} \label{ref_scalar}
\mathcal{R}_{\rm wall}&={A\over B}\left(\frac{r_{+}}{r_{+}-r_{-}}\right)^{-2i\sigma}e^{i\delta_{\rm wall}}~,
\end{align}
where, $\delta_{\rm wall}$ is the additional phase factor associated with the reflectivity of the surface of the ECO. In order to fix the phase factor $\delta_{\rm wall}$, we refer back to the case of perfectly reflective surface, i.e., the case in which $\mathcal{R}_{\rm wall}=1$. In this case from the Dirichlet boundary condition on the surface of the ECO, we obtain, $(A/B)z_{\rm wall}^{-2i\sigma}=-1$. Therefore, we choose the phase factor $\delta_{\rm wall}$, such that, $e^{i\delta_{\rm wall}}\{r_{+}/(r_{+}-r_{-})\}^{-2i\sigma}=z_{\rm wall}^{-2i\sigma}$ and hence using \ref{A_by_B} and \ref{ref_scalar}, we obtain,
\begin{align} \label{ref_scalarN}
\mathcal{R}_{\rm wall}&=\left({A\over B}\right)z_{\rm wall}^{-2i\sigma}=-\bigg[{R_{+}+i(-1)^{\ell}k^{2\ell+1}LS_{+}\over R_{-}+i(-1)^{\ell} k^{2\ell+1}LS_{-}}\bigg] z_{\rm wall}^{-2i\sigma}~.
\end{align}
Usually, it is not easy to obtain an exact analytic solution of \ref{ref_scalar}, but we can solve it approximately in the low frequency limit in order to obtain the QNM frequencies. We can assume $\sigma\ll 1$, i.e., we are considering modes having frequencies near the superradiant bound $\omega=m\Omega_{+}$, along with we are taking the low frequency limit $M\omega\ll 1$. In these limits, to leading order, $R_{+}\approx R_{-}$ and $S_{+}\approx 0\approx S_{-}$. Thus from \ref{ref_scalar}, using the tortoise coordinates we obtain,
\begin{align} \label{match_low_freq}
e^{2ir_{\rm *(wall)}\tilde{\omega}}=-\mathcal{R}_{\rm wall}=e^{i(2n+1)\pi+\ln \mathcal{R}_{\rm wall}}~.
\end{align}
From this equation we can obtain the $n$th order QNM frequency to take the following form,
\begin{align}\label{asymp_qnm}
\omega_{n}\sim m\Omega_{+}+{(2n+1)\pi\over 2r_{\rm *(wall)}}-i{\ln \mathcal{R}_{\rm wall}\over 2r_{\rm *(wall)}}~, \quad n\in \mathbb{N}
\end{align}
Note that the QNM frequencies depend explicitly on the reflectivity of the surface of ECO and also on the tidal charge, through the quantity $r_{\rm *(wall)}$. In particular, if we specialize to the case of Boltzmann reflectivity, where $\mathcal{R}_{\rm wall}\sim \exp(-|\tilde{\omega}|/2T_{+})$, we get the $n$th order QNM frequency to yield,
\begin{align}
\omega_{n}\sim\left(m\Omega_{+}+{(2n+1)\pi \over 2r_{\rm *(wall)}}\right)\left(1-{i\over 4r_{\rm *(wall)}T_{+}}\right)^{-1}~.
\end{align}
The above analysis provides the analytical estimation of the QNM frequencies at the zeroth level of the approximations described above. If we keep the next higher order contribution, then the QNM frequency for the $n$th mode would be given by \ref{asymp_qnm}, with $\mathcal{R}_{\rm wall}$, replaced by $\mathcal{R}_{\rm wall}R_{\rm BH}$, where $R_{\rm BH}$ is the reflectivity of the angular momentum barrier, which can be determined from \ref{AppSuper}.   

\subsection{Quasi-normal mode frequencies for electromagnetic perturbation}

For the electromagnetic perturbation we use the appropriate boundary condition given in \ref{wall_ref} for s=-1. Using \ref{A_by_B} we can write $\mathcal{R}_{wall}$ as

\begin{align} 
&\mathcal{R}_{wall}={A\over B}\left|{\bar B^2\over E}\right|^{- 1} z_{0}^{2i\sigma}
\\
&\mathcal{R}_{wall}=-{\Gamma(l+1-1)\over \Gamma(l+1+1)}
\bigg[{R_++i(-1)^l k^{2l+1}LS_+\over R_-+i(-1)^l k^{2l+1}LS_-}\bigg]\left|{\bar B^2\over E}\right|^{- 1}  z_{0}^{2i\sigma}. \label{ref_electro}
\end{align}
We can solve this in the low frequency limit to obtain the quasi-normal frequencies. We can assume $\sigma\ll 1$  the low frequency limit $M\omega\ll 1$ in \ref{ref_electro} and using the tortoise coordinates we get
\begin{align} \label{match_low_freq}
z_0^{2i \sigma}=e^{2i\sigma r_0^*(r_+-r_-)/(r_+^2+a^2)}=\mathcal{R}_{wall}.
\end{align}
Assuming the Boltzmann reflectivity of the membrane \ref{ref_b}, from this equation we can write the quasi-normal frequencies as \cite{Wang:2019rcf}
\begin{align}
\omega_{n}\sim\left(m\Omega_{+}+{(2n)\pi \over 2r_{\rm *(wall)}}\right)\left(1-{i\over 4r_{\rm *(wall)}T_{+}}\right)^{-1}~.
\end{align}
we can see that in this case the phase for the electromagnetic perturbation differ from that of the scalar case by a factor of $\pi$.

\subsection{Quasi-normal mode frequencies for gravitational perturbation}

For the gravitational perturbation we use the appropriate boundary condition given in \ref{wall_ref} for s=-2. Using \ref{A_by_B} we can write $\mathcal{R}_{wall}$ as

\begin{align}
&\mathcal{R}_{wall}={A\over B}\left|{|C|^2\over D}\right|^{- 1} z_{0}^{2i\sigma}
\\
&\mathcal{R}_{wall}=-{\Gamma(l+1-2)\over \Gamma(l+1+2)}
\bigg[{R_++i(-1)^l k^{2l+1}LS_+\over R_-+i(-1)^l k^{2l+1}LS_-}\bigg]\left|{|C|^2\over D}\right|^{- 1}  z_{0}^{2i\sigma}.  \label{ref_grav}
\end{align}
Again we can solve this in the low frequency limit to obtain the quasi-normal frequencies. We can assume $\sigma\ll 1$ near the superradiant bound $\omega=m\Omega$ , which is same as the low frequency limit $M\omega\ll 1$ in \ref{ref_grav} and use the tortoise coordinates to get
\begin{align} \label{match_low_freq}
z_0^{2i \sigma}=e^{2i\sigma r_0^*(r_+-r_-)/(r_+^2+a^2)}=-\mathcal{R}_{wall}.
\end{align}
Assuming the Boltzmann reflectivity of the membrane, from this equation we can write the quasi-normal frequencies as \cite{Wang:2019rcf}
\begin{align}
\omega_{n}\sim\left(m\Omega_{+}+{(2n+1)\pi \over 2r_{\rm *(wall)}}\right)\left(1-{i\over 4r_{\rm *(wall)}T_{+}}\right)^{-1}~.
\end{align}
we can see that the frequencies are same as the scalar case.

\section{Ringdown and echoes from exotic compact object on the brane: Numerical analysis}\label{echo_numerics}

In this section, we will present the analysis of the QNM frequencies and obtain the ringdown waveform using numerical techniques. This will compliment the analysis of the previous section, where the analytical estimation for the QNM frequencies have been obtained. The analytical computation was straightforward, when expressed in terms of the Teukolsky perturbation variables. However, the numerical analysis will require use of the Detweiler function introduced in \ref{detweiler_fn}, since the associated potential is manifestly real and short range. In order to do this analysis it is convenient to introduce the asymptotic forms of the the Detweiler function in terms of standard `in' and `up' modes\cite{Mark:2017dnq,Maggio:2019zyv}:
\begin{align} \label{in_up_sol}
X^{\rm in}&\sim
\begin{cases} 
e^{-i\tilde\omega r_{*}} \qquad \qquad \qquad \qquad   \text{for} \quad r_{*}\to -\infty
\\
A_{\rm in} e^{-i\omega r}+A_{\rm out}e^{i\omega r} \qquad \text{for} \quad r_{*}\to \infty 
\end{cases}
\\
X^{\rm up}&\sim
\begin{cases} 
B_{\rm in}e^{-i\tilde\omega r_{*}}+B_{\rm out}e^{i\tilde\omega r_{*}} & \text{for} \quad r_{*}\to -\infty
\\
e^{i\omega r} & \text{for} \quad r_{*}\to \infty 
\end{cases}
\end{align}
As evident, the mode $X^{\rm in}$ denotes ingoing wave at the horizon and both ingoing and outgoing wave at infinity, which is the reminiscent of the superradiant mode for black hole spacetime. Similarly, the mode $X^{\rm up}$ denotes outgoing wave at infinity and both ingoing and outgoing wave at the horizon, which is reminiscent of the QNM for ECO, with $B_{\rm in}=1$ and $B_{\rm out}=\mathcal{R}_{\rm wall}$. For a braneworld ECO the relevant boundary condition in terms of the Detweiler function is explicitly presented in \ref{echo_sol}, where we have an outgoing wave at infinity, while near the surface of the ECO, we have an ingoing part and an outgoing reflective part with amplitude $\mathcal{R}_{\rm wall}$. 

Our ultimate goal is to determine the ringdown signal of an ECO and for that we start by constructing the Green's function of a black hole in order to compute the ringdown modes. The Green's function, satisfying the boundary condition for a black hole ($\mathcal{R}_{wall}=0$), can be written in terms of the two linearly independent homogeneous solution of the Detweiler equation \ref{in_up_sol} as
\begin{align}\label{Green_fn_BH}
G_{\rm BH}(r_{*},r_{*}')={X^{\rm in}(r^{<}_{*})X^{\rm up}(r_{*}^{>})\over W_{\rm BH}}~,
\end{align}
where, $r_{*}^{<}\equiv \textrm{min}(r_{*},r_{*}')$ and $r_{*}^{>}=\textrm{max}(r_{*},r_{*}')$, while $W_{\rm BH}$ is the Wronskian of the independent solutions of the differential equation satisfied by the Detweiler function. In terms of the Green's function \ref{Green_fn_BH} and for a source defined as $S(r_*)$, the Fourier mode of gravitational waves seen by a distant observer is obtained as,
\begin{align}\label{infty_waveform}
X(r_{*}\rightarrow \infty)&=\int dr_{*}'G_{\rm BH}(r_{*}\rightarrow \infty,r_{*}')S(r_{*}') 
=\int dr_{*}'S(r_{*}'){X^{\rm in}(r_{*}')X^{\rm up}(r_{*}\rightarrow \infty)\over W_{\rm BH}}
\nonumber
\\
&\equiv X^{\rm up}(r_{*}\rightarrow \infty)Z_{\rm BH}^{\infty}(\omega)~,
\end{align}
and in the near region, we obtain,
\begin{align}\label{near_waveform}
X(r_{*}\rightarrow -\infty)&=\int dr_{*}'G_{\rm BH}(r_{*}\rightarrow -\infty,r_{*}')S(r_{*}') 
=\int dr_{*}'S(r_{*}'){X^{\rm in}(r_{*}\rightarrow -\infty)X^{\rm up}(r_{*}')\over W_{\rm BH}}
\\
&\equiv X^{\rm in}(r_{*}\rightarrow -\infty)Z_{\rm BH}^{\rm H}(\omega)~,
\end{align}
where, $Z_{\rm BH}^{\infty}(\omega)$ and $Z_{\rm BH}^{\rm H}(\omega)$ are the response functions of the black hole to the perturbation at infinity and at the horizon respectively. 

For an ECO, on the other hand, since the reflectivity $\mathcal{R}_{\rm wall}\neq 0$, it follows that the ingoing modes at the horizon will also contribute to the modes at infinity and thus the response function of an ECO at infinity would be different  from the black hole case. Since the black hole and the ECO satisfies the same perturbation equation, namely \ref{CD_eq}, we can construct the ECO solution in terms of the Black solution and some homogeneous solution of \ref{CD_eq} satisfying the correct boundary conditions, as given by \ref{echo_sol}, at infinity and at the surface of the ECO. Following which, we may add the following homogeneous solution, $\mathcal{K}X^{\rm up}\int_{-\infty}^{\infty}dr_{*}'~X^{\rm up}(r_{*}')S(r_{*}')W_{\rm BH}^{-1}$, where symbols have their usual meaning, to the black hole waveform, presented in \ref{infty_waveform}. This will satisfy the same outgoing boundary condition at asymptotic infinity. Thus near the surface of the ECO, we will have the following solution for the Detweiler function,
\begin{align}
X^{\rm ECO}=\left(X^{\rm in}+\mathcal{K}X^{\rm up}\right)\int_{r_{\rm *(wall)}}^{\infty}dr_{*}'~\frac{X^{\rm up}(r_{*}')S(r_{*}')}{W_{\rm BH}}~.
\end{align}
Matching the Detweiler function $X^{\rm ECO}$ in the near horizon regime, with the near-horizon asymptotic modes given in \ref{echo_sol}, we obtain $\mathcal{K}$ as,
\begin{align}\label{transfer}
\mathcal{K}=\frac{\mathcal{T}_{\rm BH}\mathcal{R}_{\rm wall}e^{-2i\tilde{\omega} r_{\rm *(wall)}}}{1-\mathcal{R}_{\rm BH}\mathcal{R}_{\rm wall}e^{-2i\tilde\omega r_{\rm *(wall)}}}~.
\end{align}
Here $\mathcal{K}$ is referred as the transfer function \cite{Mark:2017dnq,Cardoso:2019rvt}, with $\mathcal{R}_{\rm BH}$ being the reflectivity of the angular momentum barrier outside the black horizon defined in \ref{super_boundary}, while the transmissivity ($\mathcal{T}_{\rm BH}$) is the coefficient of $e^{-i\tilde{\omega}r_{*}}$ in the Detweiler function in the regime between the surface of the ECO and the angular momentum barrier. Also $\mathcal{R}_{\rm wall}$ is the reflectivity of the surface of the ECO, defined in \ref{echo_sol}. It is also possible to arrive at the same expression of the transfer function using a geometric optics approximation to determine the echo amplitude as given in\cite{Mark:2017dnq,Oshita:2020dox}. In terms of the transfer function the response of the ECO at infinity can be written down as, 
\begin{align} \label{eco_response}
Z_{\rm ECO}^{\infty}=Z_{\rm BH}^{\infty}+\mathcal{K}Z_{\rm BH}^{\rm H}~.
\end{align}
This expression for the response function of the ECO at infinity will find important application while determining the ringdown waveform in a later section. 
 
\subsection{Quasi-normal modes from the Green's function}

The QNM spectrum can also be obtained from the poles of the Green's function of the ECO. Following the analysis of \cite{Mark:2017dnq} and the discussion presented above, we can write down the Green's function associated with the perturbation of ECO as,
\begin{align} \label{ECO_greens}
G_{\rm ECO}(r_{*},r_{*}')=G_{\rm BH}(r_{*},r_{*}')+\mathcal{K}{X^{\rm up}(r_{*})X^{\rm up}(r_{*}')\over W_{\rm BH}}~,
\end{align}
where the quantity $\mathcal{K}$ has been defined in \ref{transfer}. Besides the QNM frequencies of the black hole, for ECO there are additional poles of the Green's function. These arise from the zeros of the denominator of the transfer function $\mathcal{K}$, defined in \ref{transfer}, which yields the following algebraic equation for the QNM frequency $\omega$,
\begin{align}
1-\mathcal{R}_{\rm BH}\mathcal{R}_{\rm wall}e^{-2i\tilde{\omega} r_{\rm *(wall)}}=0.
\end{align}
As evident, the solution for $\omega_n$ from this equation, coincides with the analytical results presented in \ref{qnm_scalar_sec}, in the low frequency limit. Thus the QNM frequencies obtained using the Green's function are consistent with that obtained from the explicit solution of the radial perturbation equation, depicting consistency between the two approaches.

\subsection{Ringdown waveform}

\begin{figure}[h]
\begin{center}
\includegraphics[scale=.25]{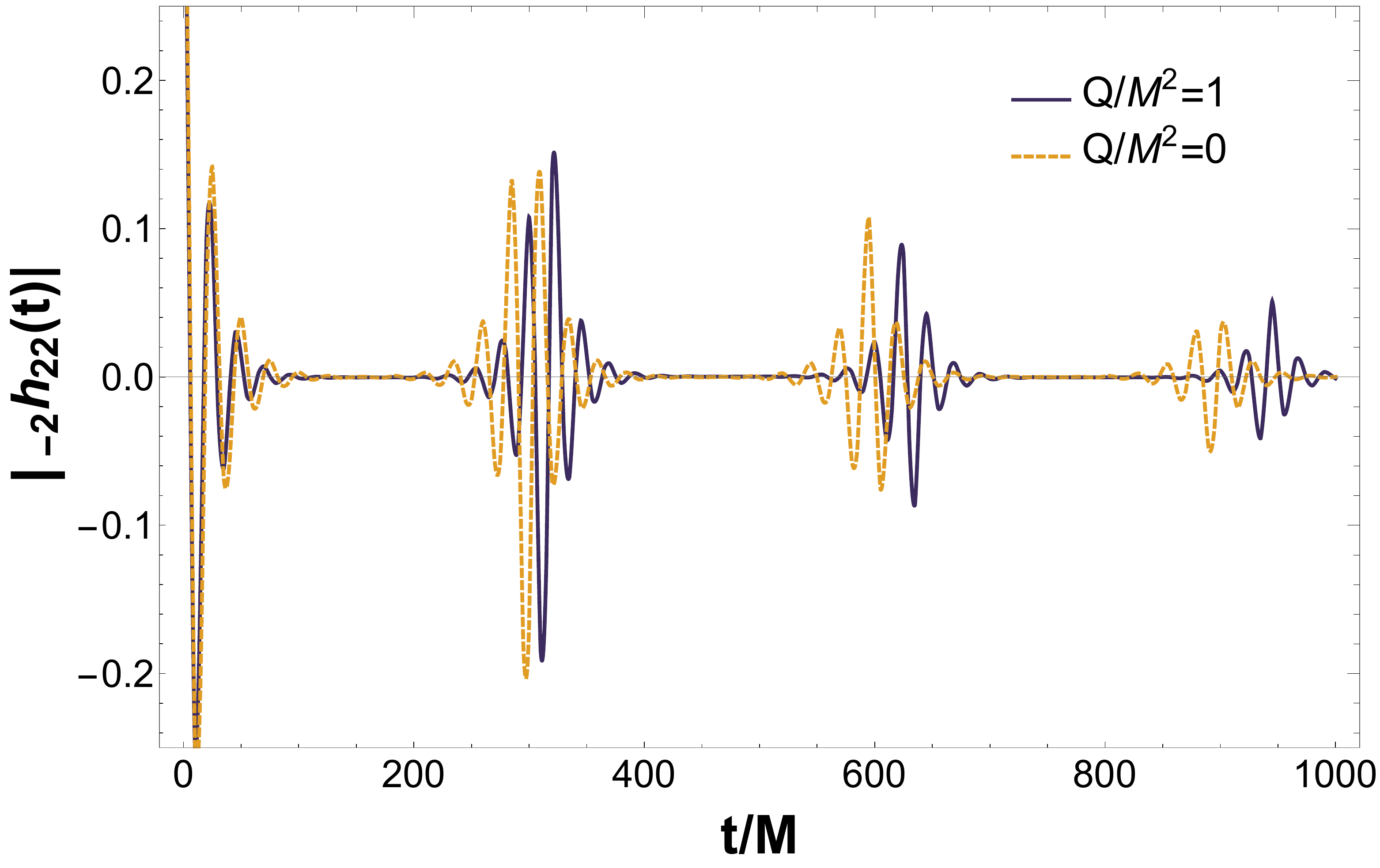}
\includegraphics[scale=.25]{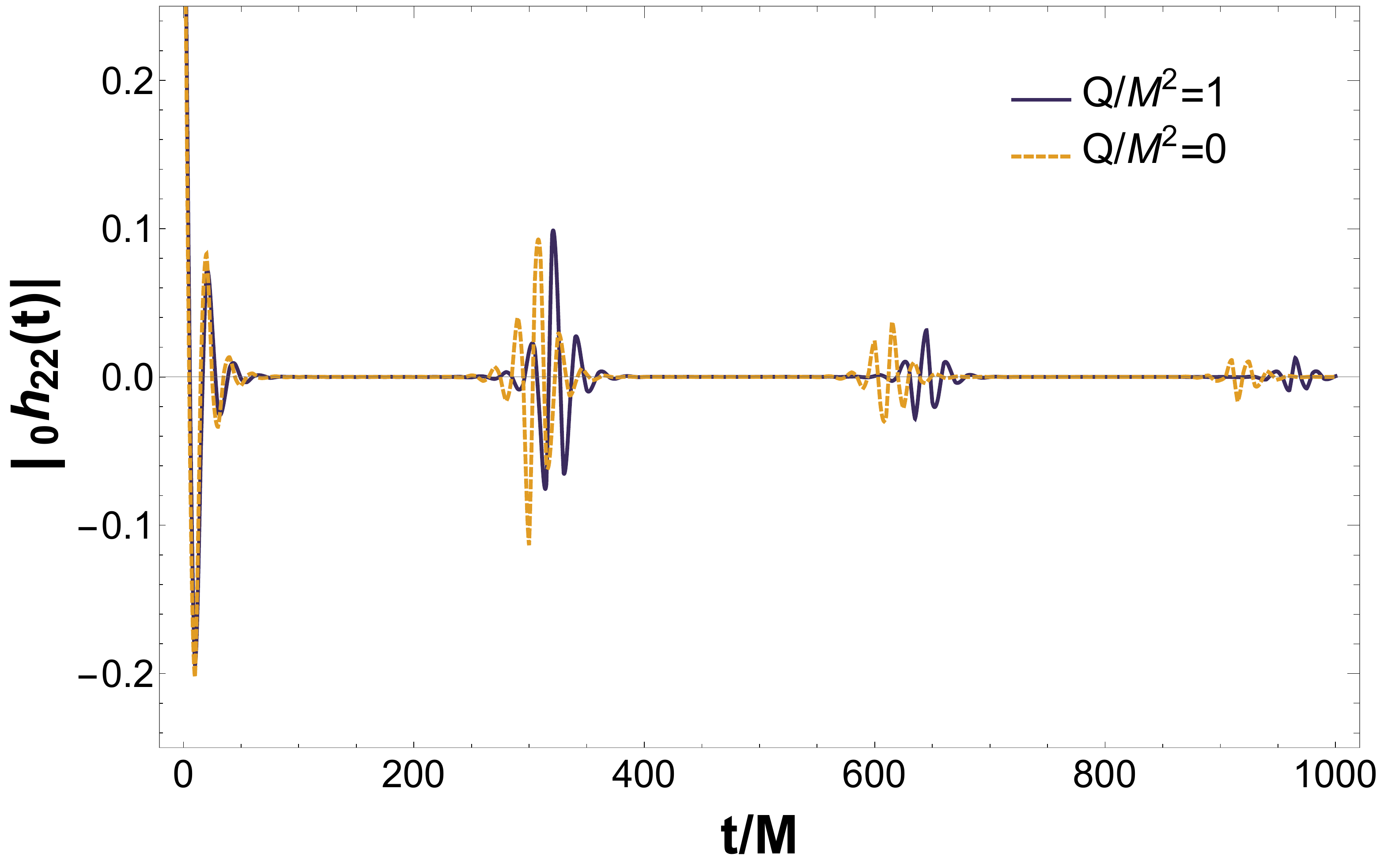}

\caption[]{The post merger time-domain ringdown signal has been presented for braneworld ECOs having dimensionless tidal charge parameters, $(Q/M^{2})=0$ and $(Q/M^{2})=1$, respectively. The above plot is for an ECO of having $a=0.67, l=2, m=2$. As evident, after the primary ringdown signal, we can see that there are additional signals considered as echo of the original signal arising out of reflection at the surface of the ECO. For non-zero values of the tidal charge parameter, these echoes are differently spaced, as the echo time-delay would depend on the tidal charge for a braneworld ECO. See text for more discussion.}\label{figure:echo}
\end{center}
\end{figure}

In this section, we will describe the derivation of the ringdown waveform for an ECO. As given in \cite{Testa:2018bzd,Maggio:2019zyv,Oshita:2020dox}, for a given source located at $r_{*}=x_{\rm s}$, such that $S=C(\omega)\delta(r_{*}-x_{\rm s})$, we can obtain a relation between $Z_{\rm BH}^{\infty}$ and $Z_{\rm BH}^{\rm H}$ as,
\begin{align} \label{ZH_relation}
Z_{\rm BH}^{\rm H}={\mathcal{R}_{\rm BH}Z_{\rm BH}^{\infty}+Z_{\rm BH}^{\infty} e^{-2i\tilde \omega x_{\rm s}}\over \mathcal{T}_{\rm BH}}~.
\end{align}
Since, the response functions $Z_{\rm BH}^{\infty}$ and $Z_{\rm BH}^{\rm H}$ have the same set of complex poles coming from the equation $W_{\rm BH}=0$, one can estimate that the near horizon response as a superposition of QNMs. Thus, using \ref{ZH_relation} we can express \ref{eco_response} in terms of quantities defined in the asymptotic limit as,
\begin{align}\label{eco_response2}
Z_{\rm ECO}^{\infty}=Z_{\rm BH}^{\infty}+\mathcal{K}\left[{\mathcal{R}_{\rm BH}Z_{\rm BH}^{\infty}+Z_{\rm BH}^{\infty}e^{-2i\tilde \omega x_{\rm s}}\over \mathcal{T}_{\rm BH}}\right]=Z_{\rm BH}^{\infty}\left({1\over 1-\mathcal{R}_{\rm BH}\mathcal{R}_{\rm wall}e^{-2i\tilde{\omega} r_{\rm *(wall)}}}\right)~,
\end{align}
where, in the last step we have used the explicit expression of the transfer function from \ref{transfer} and neglected the term coming from the product of the transfer function and the second term in the numerator of the above equation. The initial black hole ringdown amplitude $Z_{\rm BH}^{\infty}$ can be modelled based on \cite{Berti:2005ys,Maggio:2019zyv,Oshita:2020dox} for the rotating ECO in the braneworld scenario, where we avoid the details of this black hole ringdown model as our main purpose of this section is to show quantitatively how the tidal charge $Q$ would affect the ringdown waveform of a braneworld ECO and change the echo time. Thus from \ref{eco_response2}, the time domain ringdown waveform of the ECO can be computed as
\begin{align} \label{time_domain_signal}
{}_sh_{lm}(t)={1\over 2\pi}\int_{-\infty}^{\infty} d\omega~Z_{\rm ECO}^{\infty}(\omega)e^{-i\omega t}~.
\end{align}
Here $\omega$ corresponds to QNMs associated with the spin $s$ perturbation of the braneworld ECOs and $Z^{\infty}_{\rm ECO}$ is the response function of the ECO at infinity, which can be expressed in terms of black hole parameters through \ref{eco_response2}. Using the reflectivity $\mathcal{R}_{\rm BH}$ and transmissivity $\mathcal{T}_{\rm BH}$ of the classical black hole potential and choosing an appropriate initial condition, one can obtain the time domain waveform which has been plotted in \ref{figure:echo}. We can explicitly observe the distinctive echoes in the ringdown waveform, due to the presence of a reflective surface.

The time gap between two consecutive echos can be determined in terms of the black hole parameters and the location of the reflective surface. In the context of the braneworld scenario, the location of the reflective surface, given by \ref{rp_shift}, depends on the AdS length scale and hence can be estimated if the echoes are observed in the ringdown waveform of the gravitational waves. The presence of the tidal charge modifies the time gap between successive echoes, as evident from \ref{figure:echo}. Thus a measurement of the time gap (also referred to as the time delay compared to the primary QNM waveform) will enable one to determine the tidal charge parameter and hence the AdS length scale. This time delay, often referred to as the `echo time' $\Delta t_{\rm echo}$, can be expressed as twice the distance between the surface of the ECO and the maxima of the angular momentum barrier in terms of the tortoise coordinate, which reads,
\begin{align}\label{time_delay}
\Delta t_{\rm echo}= 2\left(r_{\rm *(barrier)}-r_{\rm *(wall)}\right)=2\int_{r_{\rm wall}}^{r_{\rm barrier}}{r^2+a^2\over r^2-2Mr+a^2-Q}~dr
\simeq {2(r_{+}^{2}+a^{2})\over r_{+}-r_{-}}\ln\left({r_{+}\over L}\right)~.
\end{align}
As evident, the echo time delay $\Delta t_{\rm echo}$ depends on the tidal charge parameter $Q$ through the pre-factor of the logarithm, as well as through the term $r_{+}$ within the logarithm. In addition, it also depends explicitly on the AdS length scale $L$, though only logarithmically. 
\begin{figure}[h]
\begin{center}
\includegraphics[width=11cm]{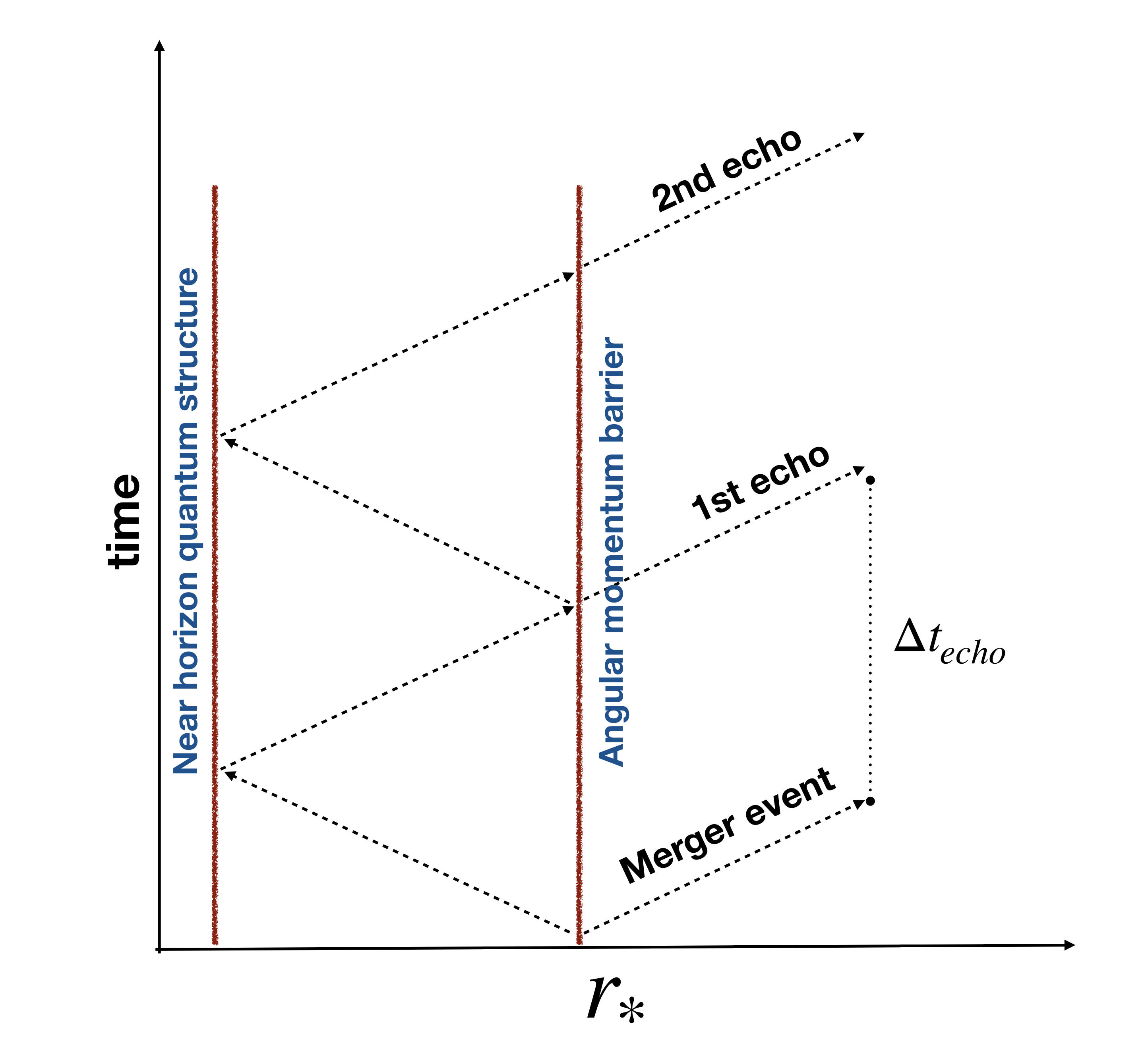}
 \caption[]{The basic structure of an ECO has been presented with a quantum structure in the near-horizon region which would partially reflect the ingoing waves incident on the merged ECO. These reflected waves would partially transmit through the angular momentum barrier and reach a distant observer as an echo of the primary signal with a time delay $\Delta t_{\rm echo}$. The echo time-delay is determined by the position of the reflective membrane placed in front of the would be horizon and the maxima of the angular momentum barrier. See text for further discussions.}\label{figure:eco_wall}
\end{center}
\end{figure}

The explicit dependence of the echo time delay on the size of the extra dimension, through the tidal charge parameter $Q$, takes the following form,
\begin{align}
\frac{\Delta t_{\rm echo}}{M}=\frac{2\left(1+\sqrt{1+\frac{Q}{M^{2}}}+\frac{Q}{2M^{2}}\right)}{\sqrt{1+\frac{Q}{M^{2}}}}\ln \left({M+\sqrt{M^{2}+Q}\over L}\right)~.
\end{align}
As evident from the above expression, presence of the extra dimension, reflected by a non-zero value of the tidal charge parameter $Q$, increases the echo time delay $\Delta t_{\rm echo}$ and hence provides a distinct signature for the existence of extra dimension. This can also be understood more intuitively by looking at \ref{figure:eco_wall}, where the echo time-delay $\Delta t_{\rm echo}$, has been explicitly depicted to depend on the size of the cavity formed by the surface of the ECO placed at $r_{\rm *(wall)}$ and the maxima of the angular momentum barrier. As obtained in \ref{epsilon}, the position of the wall is sensitive to the value of $Q$, as well as the maxima of the angular momentum barrier shifts to a larger value of the radial coordinate with a decrease in the length of extra dimension, which in turn increases the value of $Q$, thus increasing the size of the cavity and hence $\Delta t_{\rm echo}$ increases.

In particular, if we consider the case in which the dimensionless tidal charge parameter $(Q/M^{2})\sim 1$, then from the ringdown waveform one can determine the difference in the time delay with the $Q=0$ case, for a black hole of 70 solar mass, as $\delta (\Delta t_{\rm echo})=10^{-3}~\textrm{sec}$. Then from \ref{time_delay}, it follows that the AdS length scale is given by, $L\sim 10^{-15}r_{+}\sim 10^{-10}~\textrm{mm}$. Therefore, the extra dimensional length scale is in the sub micro-meter regime, which is consistent with the small-scale test of inverse square law of gravitation.     

In addition to the echo time delay $\Delta t_{\rm echo}$, another avenue to explore the effect of extra dimension on the gravitational wave is through the amplitude of the waveform. In particular, one may ask whether the amplitude of the repetitive waveform increases or, decreases with change in the size of the extra dimension, which is manifested as a change in the tidal charge parameter $Q$. However, as \ref{figure:echo} explicitly demonstrates, the amplitude of the waveform associated with both scalar and gravitational perturbation remains almost constant even if $(Q/M^{2})$ changes by unity. Thus we may argue that the amplitude of the waveform is not much sensitive to the presence of the extra dimension, since the amplitude changes very little with a change in the tidal charge parameter $Q$. This is why we have concentrated on the time delay measurement, rather than the change in the amplitude of the waveform, in comparison to the general relativistic scenario.
\section{Discussion}\label{discussion}

In this paper, we have shown that there are compelling features of higher dimensional braneworld geometries, localized on a 3-brane, that would differ from classical black holes due to the presence of quantum effects in the near horizon region. One can account for these quantum effects by studying the RS2 braneworld scenario in the framework of AdS/CFT duality \cite{Emparan_2000,Emparan:2002px,Figueras_2011}. In addition, the extension of this brane-localized solution to the bulk geometry also transforms the brane event horizon to a bulk apparent horizon. Based on these arguments we advocated that the classical black hole horizon must be either removed or modified and thus the boundary conditions imposed at the black hole horizon would change accordingly. Ideally one should be able to derive these modified near horizon boundary conditions from specific models of quantum gravity, but that is far beyond the scope of our quantitative analysis. Following which, we have assumed that there is a partially reflective membrane present in front of the would-be-horizon and the reflectivity of this membrane can either be a constant or depend on the frequency (as should be the case for a quantum black hole based on \cite{Oshita:2019sat,Dey:2020wzm}). Thus braneworld geometries must be considered as ECOs, rather than black holes with partially reflective boundary condition in the near horizon regime.

Usually, rotating horizonless ECOs are infested with the problem of superradiant instability. As shown in \cite{Cardoso:2008kj} for the scalar case and in \cite{Maggio:2018ivz} for both the case of electromagnetic and gravitational perturbations, the instability can be suppressed by tuning the reflectivity of the membrane to a value less than unity. By extending their analysis to the brane localized black holes we have found that the superradiant instability is further curbed based on the tidal charge $Q$ originating from the braneworld ECO. The tidal charge parameter is related to the size of the extra dimension, as well as the size of the horizon, as one extends the brane localised black hole solution to the bulk spacetime. In particular, the superradiant bound on the frequency decreases, as the size of the extra dimension becomes smaller (or, equivalently as the tidal charge parameter increases), thereby less number of perturbation modes experience the instability. To provide a direct estimate, based on an analytic and numerical analysis, we found that for $Q=0$ and $(a/M)=0.998$, one must allow for at least $36\%$ absorption by the membrane, in order to suppress the superradiant instability. On the other hand, for $(Q/M^{2})=1$, if the reflective surface can absorb $28\%$ of the ingoing radiation for the same $(a/M)$ ratio, the superradiant instability can be avoided. Thus non-zero tidal charge (or, presence of extra spatial dimension) indeed helps in curbing the superradiant instability for rotating ECOs. Since, for an ECO, the zero frequency modes are primarily associated with the onset of the instability, we have studied the static (zero frequency) modes and established a critical value of the black hole spin parameter ($a_{\rm{crit}}$) at which the instability appears. Our findings suggest that this critical value of the spin parameter depends on the tidal charge of the black hole in such a way that $a_{\rm{crit}}$ increases with increasing $Q$, i.e., as the size of the extra dimension decreases. This analysis further confirms our findings that as the size of the extra dimension decreases, the value of the tidal charge increases and hence the superradiant instability is suppressed for generic spin $s$ perturbations. Physically, as the size of the extra dimension decreases, the energy scale of the bulk spacetime increases. As a consequence the effect of the bulk spacetime on the brane is enhanced, resulting into a stronger gravitational field, thereby reducing the instability. The same holds true for the static case as well.

Subsequently, we have also studied plausible observational signatures of these quantum corrected braneworld ECOs by studying the associated QNMs. To start with, we have determined the QNMs of the braneworld ECOs using an analytic method in the low frequency approximation and have found that the characteristic frequencies of the ECOs differ considerably from that of the black holes as expected. The imaginary part of the frequencies show a power law decay contrary to an exponential decay for the case of a classical black hole, when we model the reflectivity of the membrane based on the Boltzmann reflectivity \cite{Wang:2019rcf}. The real part of the frequency is discrete and sensitive to the compactness of the ECO (or, the position of the reflective membrane) and the tidal charge parameter. In order to analyse the time-domain ringdown waveform, we have used a ringdown template designed to capture the non-trivial near horizon geometry of an ECO through the method of transfer function. Based on this ringdown template our study suggests that the echoes in the post-merger ringdown waveform is dependent on the higher dimensional tidal charge parameter. In particular, the echo time-delay increases with the tidal charge parameter $Q$. This is because, larger value of the tidal charge parameter implies smaller size of the extra dimension, which translates into the existence of stronger gravitational field on the brane. As a consequence the photon sphere, i.e., the maxima of the potential barrier is shifted to a larger radial distance, thereby increasing the time of flight between the surface of the ECO and the potential barrier, leading to an enlarged time delay. The enhanced time delay may allow us to put constraints on the AdS length scale as and when such echo waveforms are observed. As we have explicitly demonstrated above, for the value of the tidal charge parameter considered here, the associated AdS length scale turns out to be sub micro-meter, consistent with the small scale tests of inverse square law. 

There are several future avenues to explore in these directions. For example, effect of the reflective membrane on the polarization of the gravitational waves have not been considered so far. This becomes more important in the braneworld scenario, since the number of polarization modes may increase, leading to certain observable consequences. In addition, we have not discussed about other observable consequences of braneworld ECOs, e.g., the effect of the reflective boundary condition on the tidal Love number, the QNM spectrum in the context of black string braneworld, the phenomenon of tidal heating among many others. Furthermore, the theoretical model of the braneworld ECOs presented here, is in the semi-classical regime and involves the effect of the back-reaction to the CFT of the leading order. Better insights into the theoretical models can be obtained if the AdS/CFT dictionary can be used in a more concrete manner, so that more satisfactory prediction regarding the underlying quantum gravity model can be obtained. We hope to address these exciting issues in future works.   

\section*{Acknowledgements}

Research of S.C. is funded by the INSPIRE Faculty fellowship from DST, Government of India (Reg. No. DST/INSPIRE/04/2018/000893) and by the Start-Up Research Grant from SERB, DST, Government of India (Reg. No. SRG/2020/000409).

\appendix
\labelformat{section}{Appendix #1} 
\labelformat{subsection}{Appendix #1}
\section{Detweiler transformation}\label{AppDet}

We have defined the Detweiler function  in \ref{detweiler_fn} and in this appendix we will outline how to derive \ref{CD_eq} based on some constraints on $\alpha(r)$ and $\beta(r)$. From \ref{detweiler_fn} the derivative of the Detweiler function with respect to r is given as 
\begin{align}\label{X1}
&\frac{d{}_sX_{lm}}{dr}=\Delta^{s/2-1}G(r^{2}+a^{2})^{\frac{3}{2}}\left[\alpha \,{}_sR_{lm}+\beta\Delta^{s+1}{}_sR_{lm}^{'}\right]
\nonumber\\
&+\Delta^{s/2}\sqrt{r^{2}+a^{2}}\left[\alpha^{'}\,{}_sR_{lm}+{}_sR_{lm}^{'}\alpha+ \beta^{'}\Delta^{s+1}{}_sR_{lm}^{'}+\beta\Delta^{s}V(r,\omega)\,{}_sR_{lm}\right],
\end{align}
where $G$, $V(r,\omega)$ are defined in \ref{G} and \ref{real_potential} respectively. Using the definition of the tortoise coordinate \ref{tortoise} we can write \ref{X1} as
\begin{eqnarray}
\frac{d\,{}_sX_{lm}}{dr_{*}}=G{}_sX_{lm}+
\Delta^{\frac{s}{2}+1}({r^{2}+a^{2}})^{-\frac{1}{2}}\left[\alpha^{'}{}_sR_{lm}+{}_sR_{lm}^{'}\alpha+ \beta^{'}\Delta^{s+1}{}_sR_{lm}^{'}+\beta\Delta^{s}V(r,\omega)\, {}_sR_{lm}\right]
\end{eqnarray}
Now, in terms of the tortoise coordinate \ref{tortoise} we can write the second derivative of the Detweiler function as 
\begin{align} \label{X2}
&\frac{d^{2}{}_sX_{lm}}{dr^{2}_{*}}=\frac{\Delta}{(r^{2}+a^{2})}\frac{d}{dr}\left(\frac{\Delta}{(r^{2}+a^{2})}\frac{d{}_sX_{lm}}{dr}\right)
\nonumber\\
&={}_sX_{lm}\frac{dG}{dr_{*}}+{}_sX_{lm}G^{2}+\frac{\Delta^{\frac{s}{2}+1}}{(r^{2}+a^{2})^{\frac{5}{2}}}[(s+1)\Delta^{'}(r^{2}+a^{2})][\alpha^{'}{}_sR_{lm}+{}_sR_{lm}^{'}\alpha
+ \beta^{'}\Delta^{s+1}{}_sR_{lm}^{'}+\beta\Delta^{s}V(r,\omega)\, {}_sR_{lm}]
\nonumber\\
&+\frac{\Delta}{(r^{2}+a^{2})^{2}}\Delta^{\frac{s}{2}}(r^{2}+a^{2})^{\frac{1}{2}}[V(r,\omega)(\alpha\, {}_sR_{lm}+\beta \Delta^{s+1}{}_sR_{lm}^{'})-(s+1)\alpha \Delta^{'}{}_sR_{lm}^{'}
\nonumber\\
&+\beta\Delta^{s+1}V(r,\omega)\,{}_sR_{lm}-(s+1)\beta^{'}\Delta^{s+1}\Delta^{'}{}_sR_{lm}^{'}+  \frac{2\alpha^{'}+(\beta^{'}\Delta^{s+1})^{'}}{\beta \Delta^{s}} \beta \Delta^{s+1}{}_sR_{lm}^{'}+\alpha^{''}{}_sR_{lm}\Delta+(\beta\Delta^{s}V(r,\omega))^{'}\Delta {}_sR_{lm}]
\end{align}
Using the definition of the Detweiler function \ref{detweiler_fn} and writing ${}_sR_{lm}^{'}$ in terms of ${}_sX_{lm}$ we can further write \ref{X2}
\begin{align}
\frac{d^{2}{}_sX_{lm}}{dr^{2}_{*}}&=\left[\frac{dG}{dr_{*}}+G^{2}+\frac{\Delta}{(r^{2}+a^{2})^{2}}V(r,\omega)+\frac{\Delta}{(r^{2}+a^{2})^{2}}\frac{(2\alpha^{'}+(\beta^{'}\Delta^{s+1})^{'})}{\beta \Delta^{s}}\right]{}_sX_{lm}
\nonumber\\
&+\frac{\Delta}{(r^{2}+a^{2})^{2}}\Delta^{\frac{s}{2}}(r^{2}+
a^{2})^{\frac{1}{2}}[\beta\Delta^{s+1}V(r,\omega)- \frac{(2\alpha^{'}+(\beta^{'}\Delta^{s+1})^{'})}{\beta \Delta^{s}}  \alpha +\alpha^{''}\Delta+
(\beta\Delta^{s}V(r,\omega))^{'}\Delta 
\nonumber\\
&+((s+1)\Delta^{'}(\alpha^{'}+\beta\Delta^{s}V(r,\omega)))]\,{}_sR_{lm}
\end{align}
We can write this equation in a more compact way as
\begin{eqnarray}
\frac{d^{2}{}_sX_{lm}}{dr^{2}_{*}}-V_D(r,\omega)\,{}_sX_{lm}=F\,{}_sR_{lm}
\end{eqnarray}
Where ;
\begin{equation}
V_{D}(r,\omega)\equiv\left[\frac{dG}{dr_{*}}+G^{2}+\frac{\Delta}{(r^{2}+a^{2})^{2}}V(r,\omega)+\frac{\Delta}{(r^{2}+a^{2})^{2}}\frac{(2\alpha^{'}+(\beta^{'}\Delta^{s+1})^{'})}{\beta \Delta^{s}}\right]
\end{equation}
And 
\begin{align}
F\equiv \frac{\Delta}{(r^{2}+a^{2})^{2}}\Delta^{\frac{s}{2}}(r^{2}+
a^{2})^{\frac{1}{2}}[\beta\Delta^{s+1}V(r,\omega)- \frac{(2\alpha^{'}+(\beta^{'}\Delta^{s+1})^{'})}{\beta \Delta^{s}}  \alpha +\alpha^{''}\Delta\nonumber\\+(\beta\Delta^{s}V(r,\omega))^{'}\Delta +((s+1)\Delta^{'}(\alpha^{'}+\beta\Delta^{s}V(r,\omega)))]
\end{align}
Now $\alpha$ and $\beta$ suppose to be satisfy the following relation
\begin{equation}
\alpha^{2}-\alpha^{'}\beta\Delta^{s+1}+\alpha\beta^{'}\Delta^{s+1}-\beta^{2}V(r,\omega)\Delta^{2s+1}= constant
\end{equation}
Taking a derivative with respect to r and demanding that ($\beta \Delta^{s})\neq0$ we can end up with,
\begin{eqnarray}
\alpha^{''}\Delta+\Delta(\beta\Delta^{s}V(r,\omega))-\frac{\alpha}{\beta \Delta^{s}}(2\alpha^{'}+(\beta^{'}\Delta^{s+1})^{'})+
\nonumber\\
(s+1)\Delta^{'}(\alpha\Delta^{'}+\beta\Delta^{s}V(r,\omega))+\beta^{'}\Delta^{s+1}V(r,\omega)= 0
\end{eqnarray}
Using equation(25) we find that, $F=0$; so finally we get,
\begin{equation}
\frac{d^{2}\,{}_sX_{lm}}{dr^{2}_{*}}-V_D(r,\omega){}_sX_{lm}=0
\end{equation}
So we can conclude that Detweiler function satisfies the  linear homogeneous second-order ordinary differential equation (comparable to a Schr\"odinger like eigenvalue problem).
\section{Analytic solution for scalar, electromagnetic and gravitational perturbations}\label{app_perturbation}

The perturbation equations for the radial and angular parts, given in \ref{radial_perturbation} and \ref{angular_perturbation} respectively, are usually solved using numerical techniques, otherwise, one needs to invoke certain approximations in order to solve these equations analytically. In this section we use the asymptotic solution matching technique in order to obtain an approximate solution of the radial wave equation, presented in \ref{radial_perturbation}. First of all, it is necessary to consider the low frequency approximation, defined as $M\omega \ll 1$ to simplify the perturbation equations, allowing us to solve them in terms of special functions. In addition, we divide the spacetime outside the rotating ECO described by \ref{metric} into a near-region $(r-r_{+}\ll 1/\omega)$, an asymptotic far-region $(r-r_{+}\gg M)$ and an intermediate matching region $(M\ll r-r_{+}\ll 1/\omega)$. This enables us to solve for the perturbation equation in two of the asymptotic limits and then match the solutions order by order in the overlapping matching region. In what follows, we first solve for the radial perturbation equation in the near-regime and then in the asymptotic regime, before matching in the intermediate region. 

\subsection{Solution near the surface of the ECO}

In this section, we will solve the radial perturbation equation in the region close to the surface of the ECO, which corresponds to $(r-r_{+}\ll 1/\omega)$. In which case in the low frequency limit ($M\omega \ll 1$) we obtain, $K\sim (r_{+}^{2}+a^{2})(\omega-m\Omega_{+})$. Under these approximations the radial wave equation, presented in  \ref{radial_perturbation}, takes the following form,
\begin{align}
\Delta^{1-s}\dfrac{d}{dr}\left(\Delta^{s+1}\dfrac{d{}_{s}R_{\ell m}}{dr}\right)+\Big[(r_{+}^2+a^2)^2(\omega-m\Omega_{+})^{2}-is\Delta'(r_{+}^2+a^2)(\omega-m\Omega_{+})+2isK'\Delta-\lambda\Delta \Big]{}_{s}R_{\ell m}=0~,
\end{align}
where, we have used the fact that $K'=2r\omega$ and the separation constant, in the low frequency limit can be approximated as $\lambda\sim (\ell-s)(\ell+s+1)$. It is instructive to define a new variable $z$, as in the case of static mode, as follows,
\begin{align} \label{near_cor}
z\equiv \frac{r-r_{+}}{r_{+}-r_{-}}~,
\end{align}
and in terms of this newly defined coordinate $z$, the radial perturbation equation near the surface of the ECO can be written as
\begin{align}
\left[z(z+1)\right]^{1-s}\dfrac{d}{dz}\Big\{\left[z(z+1)\right]^{1+s}\dfrac{d{}_{s}R_{\ell m}}{dz}\Big\}+\Big[\sigma^{2}+i\sigma s(1+2z)-\lambda z(z+1)\Big]{}_{s}R_{\ell m}=0
\end{align}
where, $\sigma$ is defined in terms of the frequency $\omega$ of the perturbation mode as, $\sigma\equiv(r_{+}^{2}+a^{2})(m\Omega_{+}-\omega)/(r_{+}-r_{-})$. The solution of this equation is given in terms of the hypergeometric functions ${}_{2}F_{1}$ \ref{sol_near}, which for small values of $z$ (i.e., $r-r_{+}\ll r_{+}-r_{-}$), yields
\begin{align}\label{near_small_z}
{}_{s}R_{\ell m}=Az^{-i\sigma}+Bz^{i\sigma-s}
\sim A\left(\frac{r_{+}}{r_{+}-r_{-}}\right)^{-i\sigma}e^{i\tilde{\omega}r_{*}}+B\left(\frac{r_{+}}{r_{+}-r_{-}}\right)^{i\sigma}\left(r_{+}-r_{-}\right)^{2s}\Delta^{-s}e^{-i\tilde{\omega}r_{*}}~.
\end{align}
Here, $A$ and $B$ are arbitrary constants, to be determined from the relevant boundary conditions and  $r_{*}$ is the tortoise coordinate defined in \ref{tortoise} which can be expressed in terms of $z$ in the near horizon limit, as,
\begin{align}
z\eqsim \left(\frac{r_{+}}{r_{+}-r_{-}}\right)\exp\left[\left(\frac{r_{+}-r_{-}}{r_{+}^{2}+a^{2}}\right)r_{*}\right]~.
\end{align}
The near-horizon radial perturbation, as presented in \ref{near_small_z}, can be compared with the asymptotic limits of the Teukolsky wave function presented in \ref{bound_QNMT} to yield the ratio $\mathcal{R}_{s}/\mathcal{T}_{s}$ in terms of the ratio $(A/B)$ as,
\begin{align}\label{teukolsky_reflectivity}
\dfrac{\mathcal{R}_{s}}{\mathcal{T}_{s}}=\frac{A\left(\frac{r_{+}}{r_{+}-r_{-}}\right)^{-i\sigma}}{B\left(r_+-r_-\right)^{2s}\left(\frac{r_{+}}{r_{+}-r_{-}} \right)^{i\sigma}}
=\frac{A}{B}\frac{\left(\frac{r_{+}}{r_{+}-r_{-}} \right)^{-2i\sigma}}{\left(r_+-r_-\right)^{2s}}~.
\end{align}
This will play a crucial role in the determination of the QNMs. On the other hand, in the large $r$, i.e., large $z$ limit of the near region solution, we obtain,
\begin{align}
\label{soln_near_large}
{}_{s}R_{\ell m}\sim&z^{\ell-s}~\Gamma(2\ell+1)\Bigg[\frac{A\Gamma(1-2i\sigma+s)}{\Gamma(1+\ell-2i\sigma)\Gamma(\ell+1+s)}+\frac{B\Gamma(1+2i\sigma-s)}{\Gamma(\ell+1+2i\sigma)\Gamma(\ell+1-s)}\Bigg]
\nonumber
\\
&+z^{-\ell-1-s}~{\left(-1\right)^{\ell+1+s}\over 2\Gamma(2\ell+2)}\Bigg[\frac{A\Gamma(1-2i\sigma+s)\Gamma(\ell+1-s)}{\Gamma(-\ell-2i\sigma)}+\frac{B\Gamma(1+2i\sigma-s)\Gamma(\ell+1+s)}{\Gamma(-\ell+2i\sigma)}\Bigg]~,
\end{align}
which will be important in the intermediate region, where this solution will be matched with the solution in the far region. 

\subsection{Solution in the far-region}

In the far region, i.e., with the condition $(r-r_{+})\gg M$, the radial perturbation equation reduces to the wave equation for a field of spin s having frequency $\omega$ and angular momentum $\ell$ in the flat background (since the effects of the black hole can be neglected in this region). Using the coordinate $z$ defined in \ref{near_cor}, the radial perturbation equation in the far region, can be written down as,
\begin{align}
{d^2 {}_{s}R_{\ell m}\over dz^2}+{2(1+s)\over z}{d {}_{s}R_{\ell m}\over dz}+\bigg(k^2+{2isk \over z}-{\lambda \over z^2} \bigg){}_{s}R_{\ell m}=0
\end{align}
where, we have defined, $k\equiv \omega(r_{+}-r_{-})$. The resulting solution is in terms of the confluent Hypergeometric functions \ref{infty_soln}, which for large values of $z$, takes the following form,
\begin{align} \label{far_sol_large_r}
 R^{(s)}_{lm}&\sim {1\over \omega}\Bigg[\alpha k^{s-\ell}{(-2i)^{s-\ell-1}\Gamma(2\ell+2)\over \Gamma (\ell+s+1)}+\beta k^{\ell+1+s}{(-2i)^{\ell+s}\Gamma(-2\ell)\over \Gamma (-\ell+s)}\Bigg]{e^{-i\omega r}\over r}
\nonumber
\\
&\hskip 1 cm +{1\over \omega^{2s+1}}\Bigg[\alpha k^{s-\ell}{(-2i)^{-s-\ell-1}\Gamma(2\ell+2)\over \Gamma (\ell-s+1)}+\beta k^{\ell+1+s}{(2i)^{\ell-s}\Gamma(-2\ell)\over \Gamma (-\ell-s)}\Bigg]{e^{i\omega r}\over r^{2s+1}}
\end{align}
where, $\alpha$ and $\beta$ are arbitrary constants, to be determined using appropriate boundary conditions. The boundary condition  at asymptotic infinity (stated in terms of absence of ingoing waves from infinity) suggests that the coefficient of $(1/r)e^{-i\omega r}$ must vanish identically, yielding the ratio $(\beta/\alpha)$. Using which and expanding the far-region solution for  $kz\ll 1$, we obtain,
\begin{align}\label{far_small_r}
{}_{s}R_{\ell m}\sim \alpha{(-1)^{\ell-s}\over 2}\frac{\Gamma(\ell+1+s)}{\Gamma(2\ell+2)}r^{\ell-s}+\alpha(2i\omega)^{-1-2\ell}{\Gamma(2\ell+1)\over \Gamma(1+\ell-s)}r^{-1-\ell-s}~.
\end{align}
This provides the relevant limit for the intermediate region, arising out of the far region solution. Matching of this solution with the one presented in \ref{soln_near_large}, we immediately obtain the ratio $A/B$ in terms of spin of the perturbation, angular momentum, frequency and hairs of the black hole. Further details can be found in the main text. 

\section{Analytical computation of the amplification factor associated with superradiance of braneworld ECO}

\label{AppSuper}

In \ref{superradiance} we have defined the amplification factor ${}_{s}Z_{\ell m}$ for the modes scattered from the angular momentum barrier due to the existence of negative energy states in the ergo-region for the perturbation of arbitrary spin $s$. In order to determine the amplification factor, we will follow \cite{Page:1976df} and define the amplification factor as, 
\begin{align}\label{amplification_page}
{}_{s}Z_{\ell m}=\left|{\mathcal{O}_s\mathcal{O}_{-s}\over \mathcal{I}_s \mathcal{I}_{-s} }\right|-1~,
\end{align}
which is motivated from the fact that we do not have to worry about the normalisation factor arising in front of $|(\mathcal{O}_{s}/\mathcal{I}_{s})|^{2}$ for different values of the spin $s$. We will use the solution for the radial perturbation obtained in \ref{app_perturbation}, along with the method of asymptotic solution matching, in order to derive an approximate expression for the amplification factor ${}_{s}Z_{\ell m}$ in the low frequency limit. Since the phenomenon of superradiance is concerned with the amplification by the angular momentum barrier, near the photon sphere, we will consider the ingoing part of the near-region solution given in \ref{sol_near}, to obtain,
\begin{align} \label{sol_near_in}
{}_{s}R_{\ell m}=Bz^{i\sigma-s}(1+z)^{i\sigma}{}_{2}F_{1}[-\ell+2i\sigma,\ell+1+2i\sigma;1+2i\sigma-s;-z].
\end{align}
The large $z$ behaviour of this solution is given by the following expression,
\begin{align}\label{soln_near_large_in}
{}_{s}R_{\ell m}\sim B\bigg[ z^{\ell-s}~\frac{\Gamma(2\ell+1)\Gamma(1+2i\sigma-s)}{\Gamma(1+\ell+2i\sigma)\Gamma(\ell+1-s)}
+\left(-z\right)^{-\ell-1-s}~\frac{\Gamma(1+2i\sigma-s)\Gamma(\ell+1+s)}{2\Gamma(-\ell+2i\sigma)\Gamma(2\ell+2)}\bigg]
\end{align}
Matching this with the intermediate region solution arising out of \ref{infty_soln}, we obtain,
\begin{align}
\alpha=A\frac{\Gamma(2\ell+1)\Gamma(1+2i\sigma-s)}{\Gamma(1+\ell+2i\sigma)\Gamma(\ell+1-s)}~;
\quad
\beta=A\frac{\Gamma(1+2i\sigma-s)\Gamma(\ell+1+s)}{\Gamma(-\ell+2i\sigma)\Gamma(2\ell+1)}~.
\end{align}
Using these expressions for $\alpha$ and $\beta$ in \ref{far_sol_large_r} we can write down the radial perturbation as, 
\begin{align} \label{far_sol_large_r}
 R^{(s)}_{lm}&\sim {A\over \omega}\Bigg[k^{s-\ell}{(-2i)^{s-\ell-1}\Gamma(2\ell+2)\over \Gamma (\ell+s+1)}\frac{\Gamma(2\ell+1)\Gamma(1+2i\sigma-s)}{\Gamma(1+\ell+2i\sigma)\Gamma(\ell+1-s)}
\nonumber
\\
&\hskip 2 cm +k^{\ell+1+s}{(-2i)^{\ell+s}\Gamma(-2\ell)\over \Gamma (-\ell+s)}\frac{\Gamma(1+2i\sigma-s)\Gamma(\ell+1+s)}{\Gamma(-\ell+2i\sigma)\Gamma(2\ell+1)}\Bigg]{e^{-i\omega r}\over r}
\nonumber
\\
&+{A\over \omega^{2s+1}}\Bigg[k^{s-\ell}{(-2i)^{-s-\ell-1}\Gamma(2\ell+2)\over \Gamma (\ell-s+1)}\frac{\Gamma(2\ell+1)\Gamma(1+2i\sigma-s)}{\Gamma(1+\ell+2i\sigma)\Gamma(\ell+1-s)}
\nonumber
\\
&\hskip 2 cm+k^{\ell+1+s}{(2i)^{\ell-s}\Gamma(-2\ell)\over \Gamma (-\ell-s)}\frac{\Gamma(1+2i\sigma-s)\Gamma(\ell+1+s)}{\Gamma(-\ell+2i\sigma)\Gamma(2\ell+1)}\Bigg]{e^{i\omega r}\over r^{2s+1}}
\end{align}
Using these expressions and the boundary conditions presented in \ref{asymp_ampli} we obtain, $\mathcal{O}_{s}$ as the coefficient of $r^{-(2s+1)}e^{i\omega r_{*}}$ and $\mathcal{I}_{s}$ as the coefficient of $r^{-1}e^{-i\omega r_{*}}$. Therefore, using these expressions for $\mathcal{O}_{s}$ and $\mathcal{I}_{s}$, one can also determine $\mathcal{O}_{-s}$ and $\mathcal{I}_{-s}$, respectively. These when substituted in \ref{amplification_page}, yields \ref{Z_analytic}, we have used in the main text. 

\section{Absorption cross-section}
\label{abs_app}

We obtain the absorption cross section of the ECO given in \ref{metric}. This gives us a better way of understanding the superradiance of the ECO  analyse how it is effected due to the near horizon modifications in the presence of the tidal charge. We do this analysis for the scalar perturbation to understand the quantitative behaviour of the scattering cross section.
The conserved flux associated to the radial wave equation is 
\begin{align}
F=-i2\pi\big( {}_0R_{lm}^*\Delta\partial_r\,{}_0R_{lm}-{}_0R_{lm}\Delta \partial_r\,{}_0R_{lm}^*\big) .
\end{align}Using this expression for the flux we can determine the ingoing flux from infinity ($F^{in}_\infty$), which is given as 
\begin{align} \label{fluxinf}
F^{in}_\infty=2|\alpha^2+\beta^2|,
\end{align}
while the ingoing and outgoing flux at the horizon is given as
\begin{align} \label{fluxhor}
F^{in}_{r\to r_+}=4\pi \sigma(r_+-r_-)|A|^2,
\\
F^{out}_{r\to r_+}=-4\pi \sigma(r_+-r_-)|B|^2.
\end{align}
The Absorption cross section is given as 
\begin{equation} \label{Fabs}
\sigma_{abs}={F^{in}_{r\to r_+}+F^{out}_{r\to r_+}\over F^{in}_{r\to \infty}}=4\pi\sigma(r_+-r_-){1-\mathcal{R}_{wall}\over |\alpha/A|^2+|\beta/A|^2}.
\end{equation}
One can obtain ${\alpha\over A}$ and ${\beta\over B}$ by matching the solutions in the intermediate region $(1/\omega\gg r\gg M)$ and using the appropriate boundary condition as
\begin{align} \label{amplitude_rel}
{\alpha\over A}=(\omega/2)^{-l-1/2}{\Gamma(l+3/2)\Gamma(2l+1)\over (r_+-r_-)^l\Gamma(l+1)}\bigg[{\Gamma(1-2i\sigma)\over\Gamma(1-2i\sigma+l)}+\mathcal{R}_{wall}z_0^{-2i\sigma}{\Gamma(1+2i\sigma)\over\Gamma(l+1+2i\sigma)}\bigg]
\\
{\beta\over A}=(\omega/2)^{l+1/2}{\Gamma(-l+1/2)\Gamma(-2l-1)\over (r_+-r_-)^{-l-1}\Gamma(-l)}\bigg[{\Gamma(1-2i\sigma)\over\Gamma(-1-2i\sigma)}+\mathcal{R}_{wall}z_0^{-2i\sigma}{\Gamma(1+2i\sigma)\over\Gamma(-l+2i\sigma)}\bigg].
\end{align}
These expressions will be used in \ref{Fabs} to obtain the absorption cross section in the low frequency limit.

\bibliography{QNM}

\bibliographystyle{./utphys1}

\end{document}